\documentclass[
 aip,jcp,
 amsmath,amssymb,
reprint
]{revtex4-2}

\usepackage{graphicx}% Include figure files
\usepackage{dcolumn}% Align table columns on decimal point
\usepackage{bm}% bold math
\usepackage{physics} %some physics shorthand notation 
\usepackage[utf8]{inputenc}
\usepackage[T1]{fontenc}
\usepackage{etoolbox}
 \usepackage{mathptmx}
\usepackage{mathtools}
 \usepackage{hyperref}

\usepackage{tikz}
\usetikzlibrary{arrows,positioning, math} 
\tikzset{
    >=stealth',
    punkt/.style={
           rectangle,
           rounded corners,
           draw=black, very thick,
           text width=6.5em,
           minimum height=2em,
           text centered},
    pil/.style={
           ->,
           thick,
           shorten <=2pt,
           shorten >=2pt,}
}

\newcommand{\sirandkernel}{S4 A }
\newcommand{\sireg}{S4 }
\newcommand{\simatexp}{S3 }
\newcommand{\siiomkgn}{S5 }
\newcommand{\sijust}{S2 }
\usepackage[normalem]{ulem}

\newcommand{\rev}[1]{} 
\newcommand{\add}[1]{#1} 

\makeatletter
\def\@email#1#2{
 \endgroup
 \patchcmd{\titleblock@produce}
  {\frontmatter@RRAPformat}
  {\frontmatter@RRAPformat{\produce@RRAP{*#1\href{mailto:#2}{#2}}}\frontmatter@RRAPformat}
  {}{}
}
\makeatother

\begin{document}

\title{A Gauss-Newton method for iterative optimization of memory kernels for generalized Langevin thermostats in coarse-grained molecular dynamics simulations}
\author{Viktor Klippenstein}
\email{klippenstein@cpc.tu-darmstadt.de}
\affiliation{Department of Chemistry, Technical University of Darmstadt, 64287 Darmstadt, Germany}
\author{Niklas Wolf}
\email{wolf@cpc.tu-darmstadt.de} 

\affiliation{Department of Chemistry, Technical University of Darmstadt, 64287 Darmstadt, Germany}
\author{Nico F. A. van der Vegt}
\email{vandervegt@cpc.tu-darmstadt.de}
\affiliation{Department of Chemistry, Technical University of Darmstadt, 64287 Darmstadt, Germany}

\date{\today}

\begin{abstract}

In molecular dynamics simulations, dynamically consistent coarse-grained (CG) models commonly use stochastic thermostats to model friction and fluctuations that are lost in a CG description. 
While Markovian, i.e., time-local, formulations of such thermostats allow for an accurate representation of diffusivities/long-time dynamics, a correct description of the dynamics on all time scales generally requires non-Markovian, i.e., non-time-local, thermostats. 
These thermostats are typically in the form of a Generalized Langevin Equation (GLE) determined by a memory kernel.
In this work, we use a Markovian embedded formulation of a position-independent GLE thermostat acting independently on each CG degree of freedom.
Extracting the memory kernel of this CG model from atomistic reference data requires several approximations. 
Therefore, this task is best understood as an inverse problem.
While our recently proposed approximate Newton scheme, Iterative Optimization of memory kernels (IOMK), allows for the iterative optimization of a memory kernel,  Markovian embedding remained potentially error-prone and computationally expensive. In this work, we present a IOMK-Gauss-Newton scheme (IOMK-GN) based on IOMK, that allows for the direct parameterization of a Markovian embedded model.

\end{abstract}

\maketitle

\section{Introduction}
The principles of statistical mechanics can be used to derive coarse-grained (CG) models for computer simulations of condensed-phase systems. Their formal derivation involves averaging over atomic degrees of freedom (DoFs) that are considered irrelevant to the problem under study, a process that leads to the definition of a many-body potential of mean force (MB-PMF).\cite{jinBottomupCoarseGrainingPrinciples2022,noidPerspectiveAdvancesChallenges2023} 
However, due to the high dimensionality of the problem, extracting the exact MB-PMF from fine-grained (FG) simulation trajectories is generally not feasible, but there are several approaches to derive close approximations based on effective pair potentials.\cite{lyubartsevCalculationEffectiveInteraction1995a, reithDerivingEffectiveMesoscale2003,shellRelativeEntropyFundamental2008,bernhardtIterativeIntegralEquation2021,jinBottomupCoarseGrainingPrinciples2022,noidPerspectiveAdvancesChallenges2023} 
With the use of effective pair potentials, coarse-grained molecular dynamics (CG-MD) simulations can reproduce several structural and thermodynamic properties in close agreement with their FG counterparts, extending the applicability of MD to larger length and time scales.\cite{noidPerspectiveAdvancesChallenges2023, noidMultiscaleCoarsegrainingMethod2008, kidderEnergeticEntropicConsiderations2021,  yuMultiscaleCoarsegrainedModel2021, guptaUglyBadGood2022} 

Since CG-MD simulations are based on Hamiltonian dynamics, they misrepresent the dynamic properties of the FG system.\cite{jinUnderstandingDynamicsCoarsegrained2023} This occurs because the eliminated DoFs inevitably contribute to friction and noise, which are not considered in CG-MD.\cite{Zwanzig2001,rudzinskiRecentProgressChemicallyspecific2019,schillingCoarsegrainedModellingOut2022,klippensteinIntroducingMemoryCoarseGrained2021}   
It is well known that neglecting these dissipative interactions results in a speed-up of dynamics\cite{tschopSimulationPolymerMelts1998,lopezComputerSimulationStudies2002,depaSpeedDynamicObservables2005, jinUnderstandingDynamicsCoarsegrained2023} and, more generally,  a mismatch of absolute and relative time scales for different molecular processes.\cite{fritzMultiscaleModelingSoft2011,johnsonComparisonFrictionParametrization2023}
To avoid this, the neglected friction and fluctuating forces can be introduced into the CG equation of motion (EoM), which then resembles the original CG-MD EoM augmented with a stochastic thermostat.
\cite{leiDirectConstructionMesoscopic2010,liIncorporationMemoryEffects2015,liComputingNonMarkovianCoarsegrained2017a,hanMesoscopicCoarsegrainedRepresentations2018,hanConstructingManybodyDissipative2021,klippensteinCrosscorrelationCorrectedFriction2021,klippensteinCrosscorrelationCorrectedFriction2022,klippensteinBottomUpInformedIteratively2023,tripathyDynamicalCoarsegrainedModels2023}
While in principle an exact CG EoM can be derived using a projection operator formalism (Mori-Zwanzig theory),\cite{Mori1965a,Zwanzig2001,schillingCoarsegrainedModellingOut2022}
in practice, certain approximations and assumptions must be made to derive tractable models.
For this reason, there are many different stochastic thermostats described in the literature, with more or less rigorous theoretical underpinnings.\cite{izvekovModelingRealDynamics2006,hijonMoriZwanzigFormalism2010,liComputingNonMarkovianCoarsegrained2017a,jungGeneralizedLangevinDynamics2018,wangDatadrivenCoarsegrainedModeling2020,klippensteinCrosscorrelationCorrectedFriction2021,hanConstructingManybodyDissipative2021}
Therefore, parameterizing a CG model directly from FG reference data requires
a good understanding of the relationship between the modeled and exact CG EoM. 
Otherwise, the quality of the resulting  CG model is difficult to predict.

Alternatively, the parameters of the stochastic thermostat can be iteratively optimized to reproduce some target properties.\cite{jungIterativeReconstructionMemory2017,jungGeneralizedLangevinDynamics2018,meyerNonMarkovianOutofequilibriumDynamics2020,hankeMathematicalAnalysisIterative2021,klippensteinCrosscorrelationCorrectedFriction2022,klippensteinBottomUpInformedIteratively2023,tripathyDynamicalCoarsegrainedModels2023}
While the optimization can be performed by brute-force sampling of the free parameters, \cite{gaoSemibottomupCoarseGraining2011,johnsonComparisonFrictionParametrization2023} efficient iterative methods based on physically motivated Jacobians can significantly reduce the computational costs. Recently, we proposed an iterative method (IOMK, Iterative Optimization of Memory Kernels) that allows for a faithful representation of the velocity auto-correlation function (VACF) by optimizing the kernel of a position-independent generalized Langevin equation (GLE).\cite{klippensteinCrosscorrelationCorrectedFriction2022,klippensteinBottomUpInformedIteratively2023,tripathyDynamicalCoarsegrainedModels2023}
Since non-Markovian models tend to be computationally costly, the IOMK method was used in conjunction with the extended phase space GLE thermostat due to Ceriotti \textit{et al.},\cite{ceriottiLangevinEquationColored2009a} which allows for an equivalent Markovian description of the system by introducing auxiliary momenta (aux-GLE).\cite{ferrarioNonMarkovianRelaxationProcess1979,marchesoniExtensionKramersTheory1983,ceriottiLangevinEquationColored2009a} 

In our previous work,\cite{klippensteinCrosscorrelationCorrectedFriction2021,klippensteinCrosscorrelationCorrectedFriction2022,klippensteinBottomUpInformedIteratively2023,tripathyDynamicalCoarsegrainedModels2023} the Markovian embedding required fitting the memory kernel with a (potentially large) set of parameters in every iteration. 
In this work, we formulate a procedure, referred to as IOMK-GN, that allows us to bypass this fitting step by using a Gauss-Newton type algorithm to optimize the parameters of the aux-GLE thermostat directly.

The remainder of this work is structured as follows: 
We summarize the non-Markovian GLE thermostat and its Markovian formulation in an extended phase space in Secs.~\ref{sec:gle} and \ref{sec:aux_gle}, respectively,
and formulate their parameterization as an inverse problem in Sec.~\ref{sec:stating_invp}.
In Secs.~\ref{sec:GN} and \ref{sec:reg}, we describe the Gauss-Newton method and introduce the regularization strategies we apply to improve the convergence and stability of the proposed IOMK-GN method. 
In Sec.~\ref{sec:algo}, we summarize the iterative procedure.
We test IOMK-GN on coarse-grained liquid ethanol and compare its performance with the original IOMK method in Sec.~\ref{sec:res_iomk_gn}. 
Finally, we discuss the main results in Sec.~\ref{sec:discussion} and conclude with a summary and outlook in Sec.~\ref{sec:summary}.

Additional strategies applied for improved stability and convergence, tests of the proposed regularization strategies, and computational details on the FG and CG simulations are described in the Appendixes.

\section{Theory}
\label{sec:theory}
\subsection{The isotropic GLE thermostat}
\label{sec:gle}
Consider a classical FG-Hamiltonian reference system of $n$ particles with positions and momenta, $\bm{r}$, $\bm{p}\in \mathbb{R}^{3n}$.
In systematic coarse-graining, a mapping scheme defines the transformations $\bm{r}\rightarrow \bm{R}$ and $\bm{p}\rightarrow \bm{P}$ where $\bm{R},\bm{P}\in \mathbb{R}^{3N}$ are the  positions and momenta of CG beads  with $N<n$.\cite{noidMultiscaleCoarsegrainingMethod2008}
We assume that the CG DoFs are defined by linear combinations of the FG DoFs, e.g., the mapping of a set of atoms onto its center of mass (CoM).
We assume all CG beads are equivalent and that the system is isotropic for notational simplicity.

As a CG EoM, we consider the GLE
\begin{equation}
    \dot{\bm{P}}_I(t) = \bm{F}_I(t) = \bm{F}_I^{C}(t) -\int_{-\infty}^{t}\dd s\,\tilde{K}(t-s)\bm{P}_I(s) +\delta\tilde{\bm{F}}_I^R(t),
    \label{eq:GLE}
\end{equation}
where $\bm{F}_I(t)$ is the total force acting on the $I$th CG bead at time $t$.
The conservative force, given by the negative gradient of a CG potential, is denoted by $\bm{F}_I^{C}(t)\equiv \bm{F}_I^{C}(\bm{R}(t))$. 
The dissipative force is represented by the convolution of the thermostat's memory kernel $\tilde{K}(t)$ and the CG bead's momentum $\bm{P}_I$ and is balanced by a random force denoted as $\delta\tilde{\bm{F}}_I^R$. 
By imposing the fluctuation-dissipation theorem (FDT)\cite{shinBrownianMotionMolecular2010,liIncorporationMemoryEffects2015,jungIterativeReconstructionMemory2017}
\begin{equation}
    \left\langle \delta\tilde{\bm{F}}_I^R(t)\delta\tilde{\bm{F}}_I^R(s) \right\rangle = 3Mk_BT\tilde{K}(t-s),
    \label{eq:fdt}
\end{equation}
the dissipative and random forces together act as a thermostat, ensuring canonical sampling. 
In Eq.~\ref{eq:fdt}, $M$ is the particle mass, $T$ is the temperature, and $k_B$ is the Boltzmann constant.

\subsection{Markovian embedded GLE thermostat}
\label{sec:aux_gle}
Non-Markovian models of the type of Eq.~\ref{eq:GLE} can be rendered Markovian in an extended phase space.
For the $I$th particle along one dimension $d\in\{x,y,z\}$, the Markovian EoM reads\cite{ceriottiLangevinEquationColored2009a}
\begin{equation}
    \begin{pmatrix}\dot{P}_{I,d}\\\dot{\bm{s}}_{I,d}\end{pmatrix} = \begin{pmatrix}F^C_{I,d}\\\bm{0}\end{pmatrix} -\bm{A}\begin{pmatrix}{P}_{I,d}\\{\bm{s}}_{I,d}\end{pmatrix}+\bm{B}\begin{pmatrix}0 \\\boldsymbol{\xi}_{I,d}\end{pmatrix}.
    \label{eq:GLE_aux}
\end{equation}
Here, $\boldsymbol{\xi}_{I,d}$ is a vector of uncorrelated random numbers with zero mean and unity variance, 
$\bm{s}_{I,d}\in \mathbb{R}^h$ is a set of $h$ auxiliary momenta, which are coupled to the CG particle's momentum via the drift matrix $\bm{A}\in \mathbb{R}^{h+1\times h+1}$, and  $\bm{B}\in \mathbb{R}^{h+1\times h+1}$ is the diffusion matrix. 
To link $\bm{A}$ in Eq.~\ref{eq:GLE_aux} to $\tilde{K}(t)$ in Eq.~\ref{eq:GLE}, we consider a drift matrix of the form,
\begin{equation}
    \bm{A} = \begin{pmatrix}
        0 &  \bm{A}_{Ps}^T\\
        -\bm{A}_{Ps} & \bm{A}_{ss}
    \end{pmatrix}
    \label{eq:A_mat2}
\end{equation}
with $\bm{A}_{Ps}\in \mathbb{R}^h$ and   $\bm{A}_{ss}\in \mathbb{R}^{h\times h}$.
By  integrating out the auxiliary variables in Eq.~\ref{eq:GLE_aux}, Eq.~\ref{eq:GLE} is recovered with the relationship\cite{ceriottiColoredNoiseThermostatsCarte2010a}
\begin{equation}
    \tilde{K}(t) =  \bm{A}_{Ps}^T e^{-|t|\bm{A}_{ss}}\bm{A}_{Ps}.
    \label{eq:K_to_A2}
\end{equation}
Canonical sampling is ensured by the FDT 
\begin{equation}
    Mk_BT\left(\bm{A}+\bm{A}^T\right) = \bm{B}\bm{B}^T,
    \label{eq:mat_fdt}
\end{equation}
which can only be fulfilled if $\bm{A}+\bm{A}^T$ is positive (semi-)definite.\cite{ceriottiColoredNoiseThermostatsCarte2010a}
This can be easily enforced by constructing $\bm{A}$ as the sum of a positive diagonal matrix and a skew-symmetric matrix, i.e., $ A_{kl}\geq 0$ for all $k = l$ and $A_{kl} = - A_{lk}$ for all $k\neq l$.

Eq.~\ref{eq:K_to_A2} does not depend on the basis in which the auxiliary momenta are represented.
Therefore, to reduce the number of parameters, one can always find a representation in which $\bm{A}_{ss}$ is diagonal with complex entries or 2-block diagonal with real entries without loss of generality.
Li \textit{et al.}\cite{liComputingNonMarkovianCoarsegrained2017a,wangImplicitsolventCoarsegrainedModeling2019} proposed to rewrite the 2-block diagonal matrix as a sum of damped oscillators (see Sec.~\sirandkernel in the supplementary material) and we employed the same strategy in our previous work.\cite{klippensteinIntroducingMemoryCoarseGrained2021,klippensteinCrosscorrelationCorrectedFriction2022,klippensteinBottomUpInformedIteratively2023,tripathyDynamicalCoarsegrainedModels2023}
Note, however, that directly fitting a sum of damped oscillators to the memory kernel only covers the whole valid parameter space of $\bm{A}_{ss}$ when real and imaginary parameters are considered (see Appendix~\ref{app:block}).
Additionally, the constraints that must be applied to fulfill the FDT Eq.~\ref{eq:mat_fdt} are not self-evident without linking the damped oscillator form to a drift matrix of a fixed structure.
In the remainder of this work, we utilize the full drift matrix, which amounts to $(h+1)(h+2)/2-1$ unique free parameters.
We found this approach to be numerically more stable and refer to Appendix~\ref{app:block} for the comparison and discussion of different structures for $\bm{A}_{ss}$.

\subsection{The inverse problem}  
\label{sec:stating_invp}
\add{While Eq.~\ref{eq:fdt} guarantees canonical sampling, it is an approximation for the FDT of the GLE considered in this work.\cite{vroylandtDerivationGeneralizedLangevin2022, schillingCoarsegrainedModellingOut2022, glatzelInterplayMemoryPotentials2022}
Consequently, the memory kernel $\tilde{K}$ obtained from the corresponding Volterra equations would not result in a match of the FG VACF.
}\cite{klippensteinIntroducingMemoryCoarseGrained2021, klippensteinCrosscorrelationCorrectedFriction2022, klippensteinBottomUpInformedIteratively2023}
\rev{We can now define the inverse problem to be solved.}
\add{Therefore, we define the following inverse problem to be solved.}

First, note that a VACF, defined as $C_{VV}(t)= \langle \bm{V}_I(0)\bm{V}_I(t)\rangle/3$ can always be associated with \rev{an} \add{the} integrated single-particle memory kernel \add{of a diffusive (single-particle) Mori GLE}, $G(t)=\int_0^t K(s)ds$, via the \add{corresponding} integrated Volterra equation\cite{kowalikMemorykernelExtractionDifferent2019, straubeRapidOnsetMolecular2020b}
\begin{equation}
       C_{VV}(t) - C_{VV}(0) = -\int_0^t \dd s\, G(t-s)C_{VV}(s).
       \label{eq:volterra}
\end{equation}
\rev{
It follows that $C_{VV}^\text{tgt}(t)$ and $G^\text{tgt}(t)$
}
\add{Given a target VACF $C_{VV}^\text{tgt}(t)$, in this work obtained from FG simulations, a corresponding integrated memory kernel $G^\text{tgt}(t)$ can be obtained via Eq.~\ref{eq:volterra}, such that $C_{VV}^\text{tgt}(t)$ and $G^\text{tgt}(t)$ } can in principle be used interchangeably as target functions.
%It follows that $C_{VV}^\text{tgt}(t)$ and $G^\text{tgt}(t)$\add{, in this work obtained from the FG simulation,} can in principle be used interchangeably as target functions.  
Therefore, given $G^\text{tgt}(t)$ we aim to find a drift matrix $\bm{A}$ that solves the non-linear least squares problem
\begin{equation}
    \min_{\bm{A}} \norm{\bm{G}-\bm{G}^{\text{tgt}}}^2  = \min_{\bm{A}} \sum_{m=1}^\mathcal{M} (G_m-G_m^{\text{tgt}})^2,
    \label{eq:iomk_gn_loss}
\end{equation}
where we introduced $\bm{G}$ as the  equidistantly sampled discrete representation of $G(t)$ with the $m$th entry $G_m = G(m\Delta t)$.
In the following, we aim to solve Eq.~\ref{eq:iomk_gn_loss} using the well established Gauss-Newton method.\cite{reinhardtNichtlineareOptimierung2013}

\subsection{The Gauss-Newton method}
\label{sec:GN}
We first introduce the framework and notations in general terms to simplify the discussion in the following sections. 
For a general non-linear least-squares problem, consider the residual vector $\bm{r}(\bm{x}): \mathbb{R}^\mathcal{N} \rightarrow \mathbb{R}^\mathcal{M}$, with $\bm{r}(\bm{x}) = \bm{f}(\bm{x})-\bm{f}^{\text{tgt}}$ representing the difference between a function $\bm{f}(\bm{x})$ of a set of parameters $\bm{x}$ and a target vector $\bm{f}^{\text{tgt}}$. 
The optimal parameters are the solution to 
\begin{equation}
    \min_{\bm{x}} \norm{\bm{r}(\bm{x})}^2.
\end{equation}
This problem can be solved iteratively using the Gauss-Newton method by starting from some initial guess $\bm{x}^0$ and successively solving
\begin{equation}
    \bm{x}^{i+1} = \bm{x}^i- (\bm{J}^{T}(\bm{x}^i)\bm{J}(\bm{x}^i))^{-1}\bm{J}^T(\bm{x}^i)\bm{r}(\bm{x}^i),
    \label{eq:newt}
\end{equation}
where $\bm{J}$ is the Jacobian of $\bm{f}$ given by 
\begin{equation}
\left(\bm{J}(\bm{x}^i)\right)_{mn} =\left. \frac{\partial f_m}{\partial x_n}\right|_{\bm{x}^i}.
\end{equation}
For brevity of notation, $\bm{J}$ refers to the Jacobian at the respective iteration, and we define $\bm{r}^i \equiv \bm{r}(\bm{x}^i)$.
Eq.~\ref{eq:newt} can be rewritten as
\begin{equation}
    \bm{D}\boldsymbol{\Delta}^{i+1} = \bm{b}^{i}
    \label{eq:lin_eq}
\end{equation}
with $\boldsymbol{\Delta}^{i+1} = \bm{x}^{i+1}-\bm{x}^{i}$, $\bm{D} = \bm{J}^{T}\bm{J} $ and $\bm{b}^i = -\bm{J}^T\bm{r}^i$. 
Note that for $\mathcal{N}=\mathcal{M}$ we can  set $\bm{D} = \bm{J}$ and $\bm{b}^i = -\bm{r}^i$, and recover Newton's root-finding algorithm. 

\subsection{Regularization}
\label{sec:reg}
Problems of the form of Eq.~\ref{eq:lin_eq} are ill-posed when $\bm{D}$ is not (numerically) invertible.
However, even when $\bm{D}$ is invertible, if the eigenvalues of  $\bm{D}$ are small, the resulting iterations may become unstable. 
This is commonly combated by approximating  Eq.~\ref{eq:lin_eq} with a regularization scheme\cite{englRegularizationInverseProblems1996} of the form
\begin{equation}
    (\bm{D}+\boldsymbol{\Lambda})\boldsymbol{\Delta}^{i+1} = \bm{b}^{i},
    \label{eq:lin_reg}
\end{equation}
stabilizing the procedure. 
The matrix $\boldsymbol{\Lambda}$ in Eq.~\ref{eq:lin_reg} is a regularization matrix, which can be chosen such that $\bm{D}+\boldsymbol{\Lambda}$ has sufficiently large positive eigenvalues. 
A common choice is the Tikhonov regularization,\cite{englRegularizationInverseProblems1996} where $\boldsymbol{\Lambda} = \alpha \bm{I}$ with $\bm{I}$ the $\mathcal{N} \times \mathcal{N}$ identity matrix and $\alpha > 0$.
This regularization scheme results in damping of contributions to $\boldsymbol{\Delta}^{i+1}$ from eigenvectors with eigenvalues $\ll\alpha$ while leaving contributions from eigenvectors with eigenvalues $\gg\alpha$ unchanged. 

Eq.~\ref{eq:lin_reg} can be rewritten as the linearized least squares problem solved in each iteration
\begin{equation}
    \min_{\boldsymbol{\Delta}^{i+1}}\norm{\bm{J}\boldsymbol{\Delta}^{i+1} -\bm{r}^i}^2 + \alpha\norm{\boldsymbol{\Delta}^{i+1}}^2.
    \label{eq:lsq_reg}
\end{equation}
In this formulation, it is evident that choosing a large value for $\alpha$ adds a quadratic bias to the absolute step size, leading to more stable iterations while reducing the convergence rate. 
The iterative process can then get stuck at a point with a small gradient far from the global optimum. 

To simultaneously get stable iterations and fast convergence, we propose the adaptive regularization matrix
\begin{equation}
    \boldsymbol{\Lambda} = \text{diag}\left(\left|\frac{\alpha b^i_1}{ x^i_1}\right|,...,\left|\frac{\alpha b^i_n}{x^i_n}\right|,..., \left|\frac{\alpha b^i_\mathcal{N}}{x^i_\mathcal{N}}\right|\right) = \left(\boldsymbol{\Lambda}^\frac{1}{2}\right)^T\boldsymbol{\Lambda}^{\frac{1}{2}}.
    \label{eq:def_rel_reg}
\end{equation}
Formulating the least squares problem for the $i$th iteration, we find
\begin{equation}
    \min_{\boldsymbol{\Delta}^{i+1}}\norm{\bm{J}\boldsymbol{\Delta}^{i+1} -\bm{r}^i}^2 +\norm{\boldsymbol{\Lambda}^\frac{1}{2}\boldsymbol{\Delta}^{i+1}}^2
    \label{eq:lin_rel_reg}.
\end{equation}
Combining Eq.~\ref{eq:def_rel_reg} with the second term in Eq.~\ref{eq:lin_rel_reg} yields
\begin{equation}
    \norm{\boldsymbol{\Lambda}^\frac{1}{2}\boldsymbol{\Delta}^{i+1}}^2  =  \sum_{n}\left| \frac{\alpha b^i_n}{x^i_n}\right|\left|\Delta_n^{i+1}\right|^2. 
    \label{eq:lin_rel_reg2}
\end{equation}
To understand the difference between the two regularization schemes, we can consider three distinct scenarios: $b_n^i\approx x_n^i$, $b_n^i \ll x_n^i$ and $b_n^i \gg x_n^i$. In the first case, both regularization schemes are equivalent as Eq.~\ref{eq:def_rel_reg} would recover the Tikhonov matrix. In the second case, the adaptive scheme penalizes large updates of the $n$th parameter less strongly compared to the Tikhonov regularization, while in the third case, the penalty is increased.

Using the Tikhonov regularization, convergence usually slows down quickly after the first few iterations as $||\bm{b}||^2$ becomes small (see Sec.~\sireg in the supplementary material for an exemplary demonstration). With Eq.~\ref{eq:def_rel_reg}, the regularization is automatically reduced near the optimal solution, allowing for faster convergence without the need to adjust $\alpha$. 
In addition, in absolute terms, the change of larger parameters is less strongly damped. This means that Eq.~\ref{eq:def_rel_reg} allows larger updates for large parameters, while small parameters are adjusted more carefully.
We further compare both regularization schemes in Appendix~\ref{app:fit} and find significantly improved performance with the adaptive scheme for 100 randomly generated memory kernels (for further elaboration, see Sec.~\sireg in the supplementary material).
Note, however, that this adaptive regularization scheme should not be applied without additional adjustments if any parameter $x_m$ can become zero, which we avoid by setting bounds on the parameter space (see Appendix~\ref{app:bounds}). 

\section{Summary of the of iterative procedure}
\label{sec:algo}

\subsection{Preparatory step}
\begin{figure}
        \centering
        \input{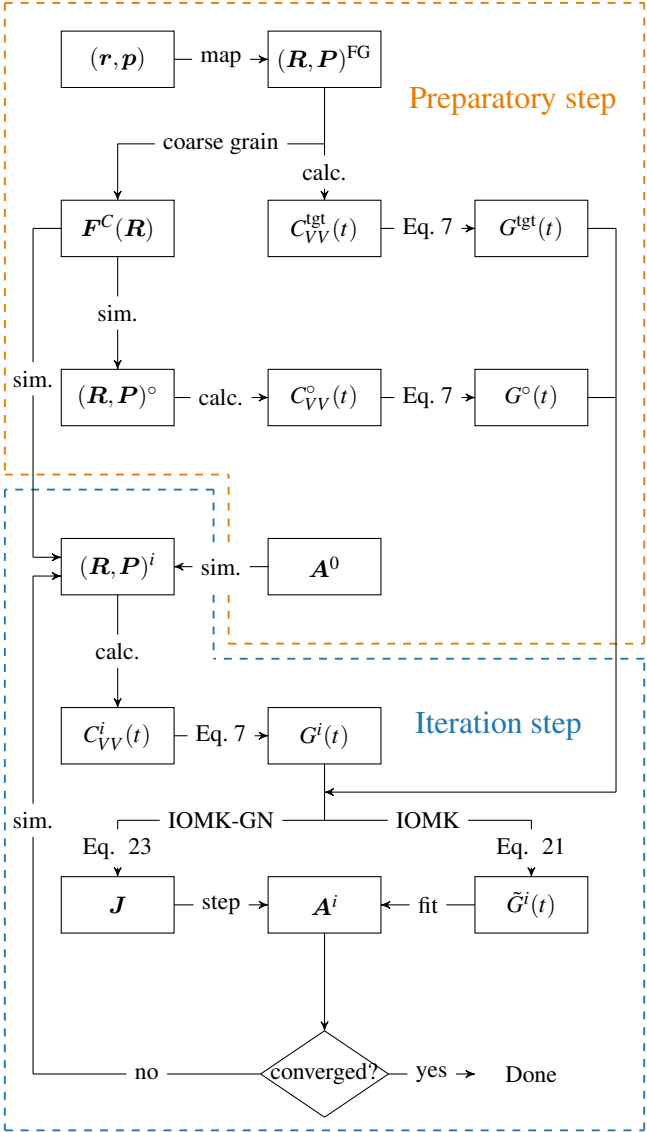}
        \caption{
        Schematic summary of the IOMK and IOMK-GN methods. 
        Each rectangle represents a quantity, each arrow represents an operation, and the diamond represents a decision.
        The procedure is generally divided into two parts:
        first, there is a preparation step in which all the properties needed for the procedure are calculated, and second, an iteration step is repeated until a convergence criterion is met or the predefined maximal number of iteration steps is reached.
        }
        \label{fig:flowchart}
\end{figure}

With the preceding discussion, we can formulate the full procedure for IOMK-GN, as summarized in the flow chart in Fig.~\ref{fig:flowchart}.
The preparatory step is equivalent in both IOMK-GN and IOMK.
First, a FG reference trajectory is produced and mapped.
With the mapped trajectory, the target VACF $C_{VV}^{\text{tgt}}(t)$ and then the integrated memory kernel are calculated, by inverting Eq.~\ref{eq:volterra}, which was found to yield reliable results\cite{kowalikMemorykernelExtractionDifferent2019} and improve the numerical stability of the inversion with a predictor-corrector step\cite{straubeRapidOnsetMolecular2020b}.
Additionally, to describe the conservative interactions in the CG system, a CG potential is derived.
Using this potential, a (Hamiltonian) CG-MD trajectory is produced to obtain  $C_{VV}^{\circ}(t)$ and $G^\circ(t)$.
Finally, an initial guess $\bm{A}^0$ is made (see Appendix~\ref{app:guess} and Sec.~\siiomkgn in the supplementary material).
Then, for every iteration, a CG GLE simulation is carried out using $\bm{A}^i$ to obtain $G^i(t)$ from $C_{VV}^i(t)$ with Eq.~\ref{eq:volterra}.
$\bm{A}^{i+1}$ is derived using either an IOMK or IOMK-GN update as described below.

\subsubsection*{IOMK update}
For the IOMK update we set $\bm{f}:= \bm{G}$, $\bm{f}^{\text{tgt}}:=\bm{G}^{\text{tgt}}$, and $\bm{x} := \tilde{\bm{G}}$. Note that this corresponds to a Newton scheme for root finding. 
The Jacobian is based on the is based on the linear approximation\cite{klippensteinBottomUpInformedIteratively2023}
\begin{equation}
    G(t) \approx G^\circ(t) + c(t)\tilde{G}(t),\label{eq:G_lin}
\end{equation}
where $c$ is a constant function of time. 
By estimating $c$ with the memory kernel from the previous iteration we arrive at
 \begin{equation}
     G^{i+1}(t) \approx G^{\circ}(t) + \frac{G^i(t)-G^\circ(t)}{\tilde{G}^{i}(t)}\tilde{G}^{i+1}(t).
     \label{eq:G_lin1}
 \end{equation}
Rewriting Eq.~\ref{eq:G_lin1} in vector notation, we obtain an approximate Jacobian as
\begin{equation}
    J_{mn} = \left.\frac{\partial G_{n}}{\partial \tilde{G}_{m} }\right|_{\tilde{\bm{G}}^i} \approx
    \left\{
    \begin{array}{cl}
     \frac{G^i_{n}- G^\circ_n}{\tilde{G}^i_{n}} &\text{ for $n = m$}\\
    0 &\text{ for $n\neq m$}
    \end{array}
    \right.
    , \label{eq:iomk_jacobian}
\end{equation}
from which follows the Newton update
\begin{equation}
    \tilde{G}^{i+1}(t) = \tilde{G}^{i}(t) - \frac{\tilde{G}^{i}(t)}{G^i(t)-G^\circ(t)}\left(G^i(t)-G^{tgt}(t)\right)
\label{eq:iomk_step}
\end{equation}
as described in our previous work.\cite{klippensteinBottomUpInformedIteratively2023}
We briefly demonstrate and discuss this approximate linearity in Sec.~\sijust in the supplementary material.
To obtain $\bm{A}^{i+1}$, we fit the integrated form of Eq.~\ref{eq:K_to_A2} given by
\begin{equation}
    \tilde{G}(t)  = \bm{A}_{Ps}^T\bm{A}_{ss}^{-1}\bm{A}_{Ps}-\bm{A}_{Ps}^T \bm{A}_{ss}^{-1}e^{-|t|\bm{A}_{ss}}\bm{A}_{Ps}
    \label{eq:G_to_A}
\end{equation}
to  $\tilde{G}^{i+1}(t)$. For that we use the Gauss-Newton method as described in Sec.~\ref{sec:GN} using the adaptive regularization scheme with $\alpha = 1$ and the additional strategies described in  Appendix ~\ref{app:GN-fit}.

\subsubsection*{IOMK-GN update}
For the IOMK-GN update, we set $\bm{f}:= \bm{G}$, $\bm{f}^{\text{tgt}}:=\bm{G}^{\text{tgt}}$, and $\bm{x} := \text{all distinct entries in } \bm{A}$, defining the Gauss-Newton update step.
With these definitions, there is no analytical relation between  $\bm{f}^{i}$ and $\bm{x}^i$, and the evaluation of  
the Jacobian would involve at least one CG simulation per iteration for every parameter. 
To avoid this, we derive an approximate Jacobian from  Eq.~\ref{eq:G_lin1}, which yields
\begin{equation}
    J_{mn} = \left. \frac{\partial G_{n}}{\partial x_{m} }\right|_{\bm{A}^i} \approx \frac{G^i_{n}- G^\circ_n}{\tilde{G}^i_{n}}  \left.\frac{\partial \tilde{G}^i_n}{\partial x_m}\right|_{\bm{A}^i}, \label{eq:iomk-gn_jacobian}
\end{equation}
where $\bm{G}^i$ is calculated from the previous iteration, $\bm{G}^\circ$ is known from the preparatory step,  $\tilde{\bm{G}}^i$ is calculated from $\bm{A}^i$ with Eq.~\ref{eq:G_to_A} and using that $\left.\partial \tilde{G}^i_n/\partial x_m\right|_{\bm{A}^i}$ can  be efficiently evaluated by finite differences (see Sec.~\simatexp in the supplementary material). Eq.~\ref{eq:G_lin1} also provides an efficient way to predict $\bm{f}^{i+1}$, without carrying out simulations. This allows us to optimize the regularization parameter  $\alpha$ in every iteration following  Appendixes ~\ref{app:step}, whereby we chose the maximal considered regularization parameter to be $\alpha_{\text{max}}=1$ using the adaptive regularization scheme.
We additionally apply the strategies described in  Appendix ~\ref{app:norm}-\ref{app:bounds}. 

\section{Results} \label{sec:res_iomk_gn}
To test the developed IOMK-GN method, we coarse-grain liquid ethanol at 300 K using a single-site center of mass mapping scheme, where we derived the conservative forces using iterative Boltzmann inversion.\cite{reithDerivingEffectiveMesoscale2003} 
Details on atomistic and CG simulations and computational details on the coarse-graining procedure are described in Appendix~\ref{app:comp}.

\begin{figure*}
    \centering
    \includegraphics{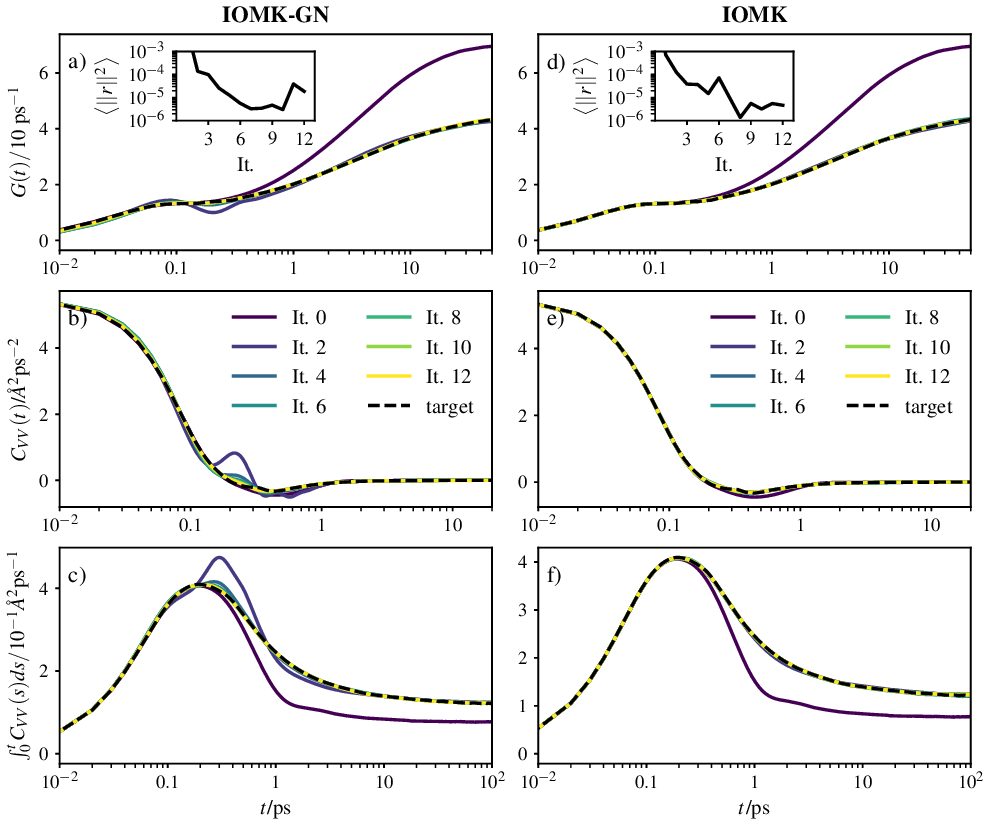}
    \caption{Summary of the results for IOMK-GN (a)-(c) and for comparison IOMK (d)-(f). The top, middle, and bottom rows show the integrated single particle memory kernel, the VACF, and the integrated VACF for a few selected iterations, compared to the respective target. The insets in (a) and (d) show the averaged squared residuals in the normalized representation.}
    \label{fig:GN_results}
\end{figure*}
Following the procedure outlined in Sec.~\ref{sec:algo}, we applied the IOMK-GN method to optimize the drift matrix for liquid CG ethanol.
Drawing on the observations discussed in Appendix ~\ref{app:fit} we use the adaptive regularization scheme with $\alpha = 1$, fitted $G^\text{tgt}(t)-G^{\circ}(t)$ to get an initial guess $\bm{A}^{0}$ and choose the dimensions of $\bm{A}$ to be $9\times 9$. 

Note that we could also use an \textit{a priori} initial guess for $\bm{A}^0$ as described in Appendix ~\ref{app:guess} (as demonstrated in Sec.~\siiomkgn in the supplementary material) but adding one fitting step before the first iteration is computationally cheap, and therefore, a simple way to derive a good initial guess. 

In Fig.~\ref{fig:GN_results} a), we compare $G^{i}(t)$ to the target $G^{\text{tgt}}(t)$ for a few selected iterations. 
We find that on short time scales, the integrated single-particle memory kernel is well reproduced by the initial guess, while on time scales beyond $\approx 1$ ps, deviations steadily increase. 
After 4 iterations, the integrated single particle memory kernel visually matches the target.  

In  Figs.~\ref{fig:GN_results} b) and c), we show the VACF and its integral for the same iterations. We note that the overall form of the VACF is well matched for the initial guess while the integrated VACF reveals the accumulated error on larger time scales, indicating too slow diffusion.
Figs.~\ref{fig:GN_results} b) and c) show that small deviations on short time scales are still present after 4 iterations and further improvement is achieved in successive iterations. Depending on the desired accuracy, the optimization can be considered converged after 4-6 iterations.

For comparison, we show the performance of the IOMK method in Figs.~\ref{fig:GN_results} d)-e). 
Hereby, we use the Gauss-Newton method for fitting, as described in Appendix \ref{app:GN-fit}, in every iteration to extract the aux-GLE parameters. 
As found in our previous work,\cite{klippensteinBottomUpInformedIteratively2023}, the IOMK method converges within 2-4 iterations. 

As a quantitative measure, the insets in Figs.~\ref{fig:GN_results} a) and d) show the respective averaged squared residual (based on the normalized representation of $G^{\text{tgt}}(t)$ as discussed in Appendix~\ref{app:norm}). After 4 iterations, both methods achieve a value smaller than $10^{-4}$, and successive iterations range between $10^{-6}$ and $10^{-4}$ for both methods.

\section{Discussion}
\label{sec:discussion} % summary
\subsection{Main findings}
Despite the fact that IOMK-GN solves an inverse problem of higher complexity than the original IOMK method, its efficiency, i.e., the number of iterations and, therefore, independent CG simulations until convergence, are comparable. 
The main improvements over our previous work\cite{klippensteinCrosscorrelationCorrectedFriction2021,klippensteinCrosscorrelationCorrectedFriction2022,klippensteinBottomUpInformedIteratively2023,tripathyDynamicalCoarsegrainedModels2023} and advantages over a naive GN method are as follows:
\begin{enumerate}
    \item The drift matrix for the aux-GLE thermostat is updated directly in each iteration using an approximation for the Jacobian. As a result, no additional fitting step is required for Markovian embedding. 
    \item We use all unique entries of the drift matrix (see Eq.~\ref{eq:G_to_A}) instead of $2\times 2$ blocks in $\bm{A}_{ss}$, as is the case for sums of damped oscillators. We find that the larger number of parameters increases numerical stability and convergence during optimization (see Appendix~\ref{app:block}).
    \item The GN update is based on several numerical improvements like normalized integrated memory kernels and an adaptive regularization scheme to, on the one hand, increase stability and, on the other hand, improve convergence.
\end{enumerate}

\subsection{Relation of IOMK and IOMK-GN to other methods}
The IOMK method\cite{klippensteinCrosscorrelationCorrectedFriction2022,klippensteinBottomUpInformedIteratively2023} was originally inspired by the iterative method for memory reconstruction as proposed by Jung \textit{et al.}.\cite{jungGeneralizedLangevinDynamics2018}
One goal in which we succeeded was to improve the convergence and, therefore, minimize the computational cost by reducing the number of iterations by one to two orders of magnitude.\cite{klippensteinCrosscorrelationCorrectedFriction2022}
Similarly, IOMK-GN was motivated by the work of Wang \textit{et al.}, who proposed using Gaussian process regression (GPR) to optimize aux-GLE parameterizations, which circumvents the problem of memory kernel fitting.\cite{wangDatadrivenCoarsegrainedModeling2020} 
While the proposed approach is intriguing, the computational cost can be high due to the need for hundreds of CG simulations. 
Here, the IOMK-GN method solves an equivalent inverse problem, again with about two orders of magnitude fewer CG simulations.

The rapid convergence of both IOMK and IOMK-GN is based on the approximate but accurate Jacobian derived from Eq.~\ref{eq:G_lin1}, while the Jacobian due to Jung \textit{et al.}\cite{jungGeneralizedLangevinDynamics2018} was originally based on the infinite mass limit,\cite{jungIterativeReconstructionMemory2017} which is often a strong approximation in molecular coarse-graining.
The  GPR method proposed by Wang~\textit{et al.}, is a derivative-free method that interpolates between sampled points in the parameter space. This has the advantage that no information about the relationship between the target function and the parameters must be available, which comes with the cost of a thorough sampling of the parameter space.

\

Therefore, the strategies presented in this work represent a significant step towards simplifying the development of dynamically consistent CG models.
Furthermore, we believe that similar ideas, as presented in this work, can be applied when parameterizing different dissipative thermostats.
Therefore, studying the effect of dissipative thermostats of different flavors on a broader range of dynamic properties could yield further relationships similar to Eq.~\ref{eq:G_lin1}.
This could, for example, allow the development of stable and efficient iterative methods for parameterizing configuration-dependent thermostats, where a fully bottom-up parameterization (e.g., via configuration-dependent time correlation functions\cite{izvekovModelingRealDynamics2006,leiDirectConstructionMesoscopic2010,liIncorporationMemoryEffects2015,deichmannBottomupApproachRepresent2018}) can be cumbersome and unreliable, even for Markovian models. \cite{trementConservativeDissipativeForce2014,deichmannBottomupDerivationConservative2014,lemarchandCoarsegrainedSimulationsCis2017} 
In some studies using dissipative particle dynamics (DPD),\cite{hoogerbruggeSimulatingMicroscopicHydrodynamic1992,grootDissipativeParticleDynamics1997} a hybrid approach has been used, in which the distance dependence of the pair friction was \textit{a priori} fixed,\cite{erikssonEffectiveThermostatInduced2008} or determined by bottom-up techniques,\cite{gaoSemibottomupCoarseGraining2011} while the remaining parameters were determined by brute-force sampling of the parameters. \cite{erikssonEffectiveThermostatInduced2008,gaoSemibottomupCoarseGraining2011} 
While this is feasible in Markovian models as long as the number of target properties and free parameters is small
systematic derivative-based optimization approaches similar to IOMK-GN should be more effective in more complex systems, such as molecular mixtures and polymer blends.

In many applications, the memory kernel can be precalculated, at least approximately, directly from FG reference data or is analytically linked to directly accessible time correlation functions.
For example, several groups have studied memory kernels of freely diffusing \cite{leiDatadrivenParameterizationGeneralized2016,liComputingNonMarkovianCoarsegrained2017a,kowalikMemorykernelExtractionDifferent2019, straubeRapidOnsetMolecular2020b,bockiusModelReductionTechniques2021} or restrained\cite{daldropExternalPotentialModifies2017} tagged particles and generalized reaction coordinates,\cite{ma2016derivation,groganDatadrivenMolecularModeling2020, ayazNonMarkovianModelingProtein2021,daltonFastProteinFolding2023} as in the dynamics of protein folding\cite{ayazNonMarkovianModelingProtein2021,daltonFastProteinFolding2023} or the dynamics of intra-molecular degrees of freedom. \cite{ma2016derivation,sheDatadrivenConstructionStochastic2023} 

In such cases, Markovian embedding amounts to fitting a Prony series to memory kernels or, by proxy, other time correlation functions, for which different strategies have been proposed,\cite{leiDatadrivenParameterizationGeneralized2016,groganDatadrivenMolecularModeling2020,wangImplicitsolventCoarsegrainedModeling2019,bockiusModelReductionTechniques2021,sheDatadrivenConstructionStochastic2023} depending on how the relationship between the memory kernel and the extended phase space EoM is formulated.
The drift matrix parameters were determined (quasi-)analytically,\cite{leiDatadrivenParameterizationGeneralized2016,ma2016derivation} by applying the Prony method\cite{bockiusModelReductionTechniques2021} or applying rational approximations for the memory kernel,\cite{leiDatadrivenParameterizationGeneralized2016,ma2016derivation}
by extracting non-Markovian features from time-correlations functions\cite{sheDatadrivenConstructionStochastic2023} or
by fitting a precalculated memory kernel using some fitting algorithm.\cite{liComputingNonMarkovianCoarsegrained2017a,yoshimotoConstructionNonMarkovianCoarsegrained2017,wangImplicitsolventCoarsegrainedModeling2019}

For example, the drift matrix can be constructed by fitting a sum of damped oscillators to the memory kernel as proposed in Refs.~\citenum{liComputingNonMarkovianCoarsegrained2017a,yoshimotoConstructionNonMarkovianCoarsegrained2017,wangImplicitsolventCoarsegrainedModeling2019}.
As in our previous work,\cite{klippensteinCrosscorrelationCorrectedFriction2021,klippensteinCrosscorrelationCorrectedFriction2022,klippensteinBottomUpInformedIteratively2023,tripathyDynamicalCoarsegrainedModels2023} the fitting procedure was not the main focus and therefore not described in detail.
Theoretically, the choice of fitting strategy does not affect the quality of the final model as long as an accurate fit is achieved, and any standard least squares solver (such as the implementations in the optimize library of SciPy\cite{virtanenSciPyFundamentalAlgorithms2020}) can be used for this purpose.
However, unstable and inefficient methods, or methods that require manual tuning (e.g., of the initial guess or bounds) to work reliably, can be a major technical hurdle to general applicability.
While IOMK-GN, due to the definition of the approximate Jacobian in Eq.~\ref{eq:iomk-gn_jacobian} allows for a direct update of the drift matrix for simulations of many interacting particles, our results indicate that the Gauss-Newton method in combination with the specialized numeric strategies can also be reliably applied for fitting precalculated memory kernels.
A direct comparison with other methods is beyond the scope of the current work, but we expect the proposed approach to be helpful for the types of applications mentioned earlier.

\section{Summary and outlook}
\label{sec:summary}
In this work, we developed  and implemented 
an efficient Gauss-Newton method (IOMK-GN) to iteratively parameterize a Markovian-embedded GLE for CG models.
We have shown that the approximate linearity between the single-particle integrated memory kernel and the integrated thermostat memory kernel in many-body particle-based CG simulations (Eq.~\ref{eq:G_lin}) can be exploited to define a cost-effective way to estimate the Jacobian of the single-particle integrated memory kernel with respect to the drift matrix parameters for the aux-GLE thermostat. 
In the context of systematic coarse-graining, IOMK-GN can be used to derive CG models that are dynamically consistent with atomistic models on all time scales within a few iterations.
During the development of IOMK-GN, we solved several numerical challenges related to Markovian embedding by least squares minimization, resulting in an efficient implementation that iteratively solves the Markovian embedding problem.
Therefore, in this work, we additionally solved all major numerical problems associated with using IOMK with Markovian-embedded GLEs.
In terms of convergence and stability, the performance of IOMK-GN and IOMK is comparable, so choosing one over the other remains a matter of preference.
The present results, combined with the results of our previous work,\cite{klippensteinCrosscorrelationCorrectedFriction2022} indicate that IOMK and IOMK-GN are orders of magnitude more efficient than previously available methods\cite{jungGeneralizedLangevinDynamics2018,wangDatadrivenCoarsegrainedModeling2020} when applied to equivalent tasks.
The implemented methods are available on GitHub  (https://github.com/vklip/d-cg).

%outlook
In this work, we have focused on the technical aspects of solving an inverse problem using an EoM without rigorously justifying its form.
Therefore, we want to emphasize that the considered GLE is approximate, and limitations in its ability to reproduce all aspects of an underlying FG system accurately are to be expected.
In particular, Eqs.~\ref{eq:GLE} and~\ref{eq:GLE_aux} neglect interparticle correlations in the memory kernel and the FDT.\cite{glatzelInterplayMemoryPotentials2022,ayazGeneralizedLangevinEquation2022,vroylandtDerivationGeneralizedLangevin2022,vroylandtPositiondependentMemoryKernel2022}
While previous work suggests that quantitatively capturing all aspects of the joint dynamics is not always possible when using configuration-independent thermostats,\cite{klippensteinBottomUpInformedIteratively2023,lyuConstructionCoarseGrainedMolecular2023} some properties, such as the distinct van Hove function, can be represented quite accurately when the single-particle dynamics are well matched.\cite{klippensteinBottomUpInformedIteratively2023,tripathyDynamicalCoarsegrainedModels2023}

With the improvements in numerical reliability and ease of use, IOMK and IOMK-GN can be more easily applied to complex systems, and the scope and limitations of the isotropic GLE thermostat in accurately representing a CG system can be more easily explored than before. 
The extension to mixtures is straightforward\cite{tripathyDynamicalCoarsegrainedModels2023}, and it should be possible to derive dynamically consistent CG models with bonded interactions and different bead types. 
In particular, when the long-time dynamics are governed by a cascade of processes relaxing on shorter time scales, as in polymer systems, it is conceivable that by targeting processes with shorter relaxation times, IOMK or IOMK-GN can be used to derive CG models with bonded interactions that accurately represent the long-time dynamics only informed by short reference trajectories. 

Of course, there are applications where a configuration-independent GLE thermostat formulation is insufficient to capture the relevant physics accurately. 
For example, the aux-GLE thermostat is not momentum conserving and thus should not be naively applied in non-equilibrium simulations.\cite{pastorinoComparisonDissipativeParticle2007} 
Dissipative particle dynamics (DPD)\cite{hoogerbruggeSimulatingMicroscopicHydrodynamic1992,grootDissipativeParticleDynamics1997} (and possibly non-Markovian DPD\cite{yoshimotoBottomupConstructionInteraction2013,liComputingNonMarkovianCoarsegrained2017a}) models are better suited for such applications.\cite{pastorinoComparisonDissipativeParticle2007}
In the context of systematic coarse-graining, direct parameterization of DPD models from FG correlation functions has been proposed,\cite{hijonMoriZwanzigFormalism2010,leiDirectConstructionMesoscopic2010} but
it is generally not easy to reliably obtain good parameterizations.\cite{trementConservativeDissipativeForce2014,
deichmannBottomupDerivationConservative2014,
lemarchandCoarsegrainedSimulationsCis2017} 
Some authors, after determining the functional form of the configuration dependent friction, optimized the absolute amplitude to match the diffusion coefficient or viscosity.\cite{erikssonEffectiveThermostatInduced2008,gaoSemibottomupCoarseGraining2011} 
This optimization is often performed by brute-force sampling of parameters, which is feasible as long as the number of free parameters and target observables is small. 
As long as a sufficiently accurate approximate Jacobian can be found, systematic Newton or Gauss-Newton methods can be explored. This would allow one to tackle more complex problems, such as mixtures or, more generally, mapping schemes with multiple bead types.

\section*{Supplementary material}
The supplementary material contains additional data and discussion on the correlation functions and memory kernels involved in the IOMK-GN method (Sec. S1), the validity of Eq.~\ref{eq:G_lin1} (Sec. S2), details on the evaluation of matrix exponentials (Sec S3), additional information on the comparison of the two regularization schemes and different tests on the effects of the strategies described in the Appendix (Sec. S4).

\begin{acknowledgments}
VK acknowledges Marvin Bernhardt for his helpful input in the initial stage of this work. NW and VK wish to acknowledge Martin Hanke-Bourgeois for fruitful discussions.
The authors acknowledge funding by the Deutsche Forschungsgemeinschaft (DFG, German Research Foundation) - Project No. 233630050 - CRC TRR 146 "Multiscale Simulation Methods for Soft Matter Systems".
\end{acknowledgments}

\section*{Author Contributions}

\textbf{Viktor Klippenstein}: Conceptualization (equal); Investigation (equal); Methodology (equal); Software (equal); Visualization
(equal); Writing - original draft (equal); Writing - review $\&$
editing (equal). \textbf{Niklas Wolf}: Conceptualization (equal); Investigation (equal); Methodology (equal); Software (equal); Visualization (equal); Writing - original draft (equal); Writing - review $\&$ editing
(equal). \textbf{Nico F. A. van der Vegt}: Conceptualization (supporting);
Funding acquisition (lead); Project administration (lead); Resources
(lead); Supervision (lead); Writing - original draft (supporting);
Writing - review $\&$ editing (equal).

\section*{Data Availability Statement}
The data that support the findings of this study are available within the article and its supplementary material. Input files for simulations and raw data of the shown figures are available at https://doi.org/10.48328/tudatalib-1414.
An implementation of IOMK-GN, IOMK and the Gauss-Newton method for fitting memory kernels is openly available on GitHub (https://github.com/vklip/d-cg).

\bibliography{IOMK-GN.bib}

%aipnum4-2.bst 2019-01-14 (MD) hand-edited version of apsrev4-1.bst
\begin{thebibliography}{78}%
\makeatletter
\providecommand \@ifxundefined [1]{%
 \@ifx{#1\undefined}
}%
\providecommand \@ifnum [1]{%
 \ifnum #1\expandafter \@firstoftwo
 \else \expandafter \@secondoftwo
 \fi
}%
\providecommand \@ifx [1]{%
 \ifx #1\expandafter \@firstoftwo
 \else \expandafter \@secondoftwo
 \fi
}%
\providecommand \natexlab [1]{#1}%
\providecommand \enquote  [1]{``#1''}%
\providecommand \bibnamefont  [1]{#1}%
\providecommand \bibfnamefont [1]{#1}%
\providecommand \citenamefont [1]{#1}%
\providecommand \href@noop [0]{\@secondoftwo}%
\providecommand \href [0]{\begingroup \@sanitize@url \@href}%
\providecommand \@href[1]{\@@startlink{#1}\@@href}%
\providecommand \@@href[1]{\endgroup#1\@@endlink}%
\providecommand \@sanitize@url [0]{\catcode `\\12\catcode `\$12\catcode
  `\&12\catcode `\#12\catcode `\^12\catcode `\_12\catcode `\%12\relax}%
\providecommand \@@startlink[1]{}%
\providecommand \@@endlink[0]{}%
\providecommand \url  [0]{\begingroup\@sanitize@url \@url }%
\providecommand \@url [1]{\endgroup\@href {#1}{\urlprefix }}%
\providecommand \urlprefix  [0]{URL }%
\providecommand \Eprint [0]{\href }%
\providecommand \doibase [0]{https://doi.org/}%
\providecommand \selectlanguage [0]{\@gobble}%
\providecommand \bibinfo  [0]{\@secondoftwo}%
\providecommand \bibfield  [0]{\@secondoftwo}%
\providecommand \translation [1]{[#1]}%
\providecommand \BibitemOpen [0]{}%
\providecommand \bibitemStop [0]{}%
\providecommand \bibitemNoStop [0]{.\EOS\space}%
\providecommand \EOS [0]{\spacefactor3000\relax}%
\providecommand \BibitemShut  [1]{\csname bibitem#1\endcsname}%
\let\auto@bib@innerbib\@empty
%</preamble>
\bibitem [{\citenamefont {Jin}\ \emph {et~al.}(2022)\citenamefont {Jin},
  \citenamefont {Pak}, \citenamefont {Durumeric}, \citenamefont {Loose},\ and\
  \citenamefont {Voth}}]{jinBottomupCoarseGrainingPrinciples2022}%
  \BibitemOpen
  \bibfield  {author} {\bibinfo {author} {\bibfnamefont {J.}~\bibnamefont
  {Jin}}, \bibinfo {author} {\bibfnamefont {A.~J.}\ \bibnamefont {Pak}},
  \bibinfo {author} {\bibfnamefont {A.~E.~P.}\ \bibnamefont {Durumeric}},
  \bibinfo {author} {\bibfnamefont {T.~D.}\ \bibnamefont {Loose}},\ and\
  \bibinfo {author} {\bibfnamefont {G.~A.}\ \bibnamefont {Voth}},\ }\bibfield
  {title} {\enquote {\bibinfo {title} {Bottom-up {{Coarse-Graining}}:
  {{Principles}} and {{Perspectives}}},}\ }\href
  {https://doi.org/10.1021/acs.jctc.2c00643} {\bibfield  {journal} {\bibinfo
  {journal} {Journal of Chemical Theory and Computation}\ }\textbf {\bibinfo
  {volume} {18}},\ \bibinfo {pages} {5759--5791} (\bibinfo {year}
  {2022})}\BibitemShut {NoStop}%
\bibitem [{\citenamefont {Noid}(2023)}]{noidPerspectiveAdvancesChallenges2023}%
  \BibitemOpen
  \bibfield  {author} {\bibinfo {author} {\bibfnamefont {W.~G.}\ \bibnamefont
  {Noid}},\ }\bibfield  {title} {\enquote {\bibinfo {title} {Perspective:
  {{Advances}}, {{Challenges}}, and {{Insight}} for {{Predictive Coarse-Grained
  Models}}},}\ }\href {https://doi.org/10.1021/acs.jpcb.2c08731} {\bibfield
  {journal} {\bibinfo  {journal} {The Journal of Physical Chemistry B}\
  }\textbf {\bibinfo {volume} {127}},\ \bibinfo {pages} {4174--4207} (\bibinfo
  {year} {2023})}\BibitemShut {NoStop}%
\bibitem [{\citenamefont {Lyubartsev}\ and\ \citenamefont
  {Laaksonen}(1995)}]{lyubartsevCalculationEffectiveInteraction1995a}%
  \BibitemOpen
  \bibfield  {author} {\bibinfo {author} {\bibfnamefont {A.~P.}\ \bibnamefont
  {Lyubartsev}}\ and\ \bibinfo {author} {\bibfnamefont {A.}~\bibnamefont
  {Laaksonen}},\ }\bibfield  {title} {\enquote {\bibinfo {title} {Calculation
  of effective interaction potentials from radial distribution functions: {{A}}
  reverse {{Monte Carlo}} approach},}\ }\href
  {https://doi.org/10.1103/PhysRevE.52.3730} {\bibfield  {journal} {\bibinfo
  {journal} {Physical Review E}\ }\textbf {\bibinfo {volume} {52}},\ \bibinfo
  {pages} {3730--3737} (\bibinfo {year} {1995})}\BibitemShut {NoStop}%
\bibitem [{\citenamefont {Reith}, \citenamefont {P{\"u}tz},\ and\ \citenamefont
  {{M{\"u}ller-Plathe}}(2003)}]{reithDerivingEffectiveMesoscale2003}%
  \BibitemOpen
  \bibfield  {author} {\bibinfo {author} {\bibfnamefont {D.}~\bibnamefont
  {Reith}}, \bibinfo {author} {\bibfnamefont {M.}~\bibnamefont {P{\"u}tz}},\
  and\ \bibinfo {author} {\bibfnamefont {F.}~\bibnamefont
  {{M{\"u}ller-Plathe}}},\ }\bibfield  {title} {\enquote {\bibinfo {title}
  {Deriving effective mesoscale potentials from atomistic simulations},}\
  }\href {https://doi.org/10.1002/jcc.10307} {\bibfield  {journal} {\bibinfo
  {journal} {Journal of Computational Chemistry}\ }\textbf {\bibinfo {volume}
  {24}},\ \bibinfo {pages} {1624--1636} (\bibinfo {year} {2003})}\BibitemShut
  {NoStop}%
\bibitem [{\citenamefont {Shell}(2008)}]{shellRelativeEntropyFundamental2008}%
  \BibitemOpen
  \bibfield  {author} {\bibinfo {author} {\bibfnamefont {M.~S.}\ \bibnamefont
  {Shell}},\ }\bibfield  {title} {\enquote {\bibinfo {title} {The relative
  entropy is fundamental to multiscale and inverse thermodynamic problems},}\
  }\href {https://doi.org/10.1063/1.2992060} {\bibfield  {journal} {\bibinfo
  {journal} {The Journal of Chemical Physics}\ }\textbf {\bibinfo {volume}
  {129}},\ \bibinfo {pages} {144108} (\bibinfo {year} {2008})}\BibitemShut
  {NoStop}%
\bibitem [{\citenamefont {Bernhardt}, \citenamefont {Hanke},\ and\
  \citenamefont {{van der
  Vegt}}(2021)}]{bernhardtIterativeIntegralEquation2021}%
  \BibitemOpen
  \bibfield  {author} {\bibinfo {author} {\bibfnamefont {M.~P.}\ \bibnamefont
  {Bernhardt}}, \bibinfo {author} {\bibfnamefont {M.}~\bibnamefont {Hanke}},\
  and\ \bibinfo {author} {\bibfnamefont {N.~F.~A.}\ \bibnamefont {{van der
  Vegt}}},\ }\bibfield  {title} {\enquote {\bibinfo {title} {Iterative integral
  equation methods for structural coarse-graining},}\ }\href
  {https://doi.org/10.1063/5.0038633} {\bibfield  {journal} {\bibinfo
  {journal} {The Journal of Chemical Physics}\ }\textbf {\bibinfo {volume}
  {154}},\ \bibinfo {pages} {084118} (\bibinfo {year} {2021})}\BibitemShut
  {NoStop}%
\bibitem [{\citenamefont {Noid}\ \emph {et~al.}(2008)\citenamefont {Noid},
  \citenamefont {Liu}, \citenamefont {Wang}, \citenamefont {Chu}, \citenamefont
  {Ayton}, \citenamefont {Izvekov}, \citenamefont {Andersen},\ and\
  \citenamefont {Voth}}]{noidMultiscaleCoarsegrainingMethod2008}%
  \BibitemOpen
  \bibfield  {author} {\bibinfo {author} {\bibfnamefont {W.~G.}\ \bibnamefont
  {Noid}}, \bibinfo {author} {\bibfnamefont {P.}~\bibnamefont {Liu}}, \bibinfo
  {author} {\bibfnamefont {Y.}~\bibnamefont {Wang}}, \bibinfo {author}
  {\bibfnamefont {J.-W.}\ \bibnamefont {Chu}}, \bibinfo {author} {\bibfnamefont
  {G.~S.}\ \bibnamefont {Ayton}}, \bibinfo {author} {\bibfnamefont
  {S.}~\bibnamefont {Izvekov}}, \bibinfo {author} {\bibfnamefont {H.~C.}\
  \bibnamefont {Andersen}},\ and\ \bibinfo {author} {\bibfnamefont {G.~A.}\
  \bibnamefont {Voth}},\ }\bibfield  {title} {\enquote {\bibinfo {title} {The
  multiscale coarse-graining method. {{II}}. {{Numerical}} implementation for
  coarse-grained molecular models},}\ }\href
  {https://doi.org/10.1063/1.2938857} {\bibfield  {journal} {\bibinfo
  {journal} {The Journal of Chemical Physics}\ }\textbf {\bibinfo {volume}
  {128}},\ \bibinfo {pages} {244115} (\bibinfo {year} {2008})}\BibitemShut
  {NoStop}%
\bibitem [{\citenamefont {Kidder}, \citenamefont {Szukalo},\ and\ \citenamefont
  {Noid}(2021)}]{kidderEnergeticEntropicConsiderations2021}%
  \BibitemOpen
  \bibfield  {author} {\bibinfo {author} {\bibfnamefont {K.~M.}\ \bibnamefont
  {Kidder}}, \bibinfo {author} {\bibfnamefont {R.~J.}\ \bibnamefont
  {Szukalo}},\ and\ \bibinfo {author} {\bibfnamefont {W.~G.}\ \bibnamefont
  {Noid}},\ }\bibfield  {title} {\enquote {\bibinfo {title} {Energetic and
  entropic considerations for coarse-graining},}\ }\href
  {https://doi.org/10.1140/epjb/s10051-021-00153-4} {\bibfield  {journal}
  {\bibinfo  {journal} {The European Physical Journal B}\ }\textbf {\bibinfo
  {volume} {94}},\ \bibinfo {pages} {153} (\bibinfo {year} {2021})}\BibitemShut
  {NoStop}%
\bibitem [{\citenamefont {Yu}\ \emph {et~al.}(2021)\citenamefont {Yu},
  \citenamefont {Pak}, \citenamefont {He}, \citenamefont {{Monje-Galvan}},
  \citenamefont {Casalino}, \citenamefont {Gaieb}, \citenamefont {Dommer},
  \citenamefont {Amaro},\ and\ \citenamefont
  {Voth}}]{yuMultiscaleCoarsegrainedModel2021}%
  \BibitemOpen
  \bibfield  {author} {\bibinfo {author} {\bibfnamefont {A.}~\bibnamefont
  {Yu}}, \bibinfo {author} {\bibfnamefont {A.~J.}\ \bibnamefont {Pak}},
  \bibinfo {author} {\bibfnamefont {P.}~\bibnamefont {He}}, \bibinfo {author}
  {\bibfnamefont {V.}~\bibnamefont {{Monje-Galvan}}}, \bibinfo {author}
  {\bibfnamefont {L.}~\bibnamefont {Casalino}}, \bibinfo {author}
  {\bibfnamefont {Z.}~\bibnamefont {Gaieb}}, \bibinfo {author} {\bibfnamefont
  {A.~C.}\ \bibnamefont {Dommer}}, \bibinfo {author} {\bibfnamefont {R.~E.}\
  \bibnamefont {Amaro}},\ and\ \bibinfo {author} {\bibfnamefont {G.~A.}\
  \bibnamefont {Voth}},\ }\bibfield  {title} {\enquote {\bibinfo {title} {A
  multiscale coarse-grained model of the {{SARS-CoV-2}} virion},}\ }\href
  {https://doi.org/10.1016/j.bpj.2020.10.048} {\bibfield  {journal} {\bibinfo
  {journal} {Biophysical Journal}\ }\textbf {\bibinfo {volume} {120}},\
  \bibinfo {pages} {1097--1104} (\bibinfo {year} {2021})}\BibitemShut {NoStop}%
\bibitem [{\citenamefont {Gupta}\ \emph {et~al.}(2022)\citenamefont {Gupta},
  \citenamefont {Sarkar}, \citenamefont {Tieleman},\ and\ \citenamefont
  {Singharoy}}]{guptaUglyBadGood2022}%
  \BibitemOpen
  \bibfield  {author} {\bibinfo {author} {\bibfnamefont {C.}~\bibnamefont
  {Gupta}}, \bibinfo {author} {\bibfnamefont {D.}~\bibnamefont {Sarkar}},
  \bibinfo {author} {\bibfnamefont {D.~P.}\ \bibnamefont {Tieleman}},\ and\
  \bibinfo {author} {\bibfnamefont {A.}~\bibnamefont {Singharoy}},\ }\bibfield
  {title} {\enquote {\bibinfo {title} {The ugly, bad, and good stories of
  large-scale biomolecular simulations},}\ }\href
  {https://doi.org/10.1016/j.sbi.2022.102338} {\bibfield  {journal} {\bibinfo
  {journal} {Current Opinion in Structural Biology}\ }\textbf {\bibinfo
  {volume} {73}},\ \bibinfo {pages} {102338} (\bibinfo {year}
  {2022})}\BibitemShut {NoStop}%
\bibitem [{\citenamefont {Jin}, \citenamefont {Schweizer},\ and\ \citenamefont
  {Voth}(2023)}]{jinUnderstandingDynamicsCoarsegrained2023}%
  \BibitemOpen
  \bibfield  {author} {\bibinfo {author} {\bibfnamefont {J.}~\bibnamefont
  {Jin}}, \bibinfo {author} {\bibfnamefont {K.~S.}\ \bibnamefont {Schweizer}},\
  and\ \bibinfo {author} {\bibfnamefont {G.~A.}\ \bibnamefont {Voth}},\
  }\bibfield  {title} {\enquote {\bibinfo {title} {Understanding dynamics in
  coarse-grained models. {{I}}. {{Universal}} excess entropy scaling
  relationship},}\ }\href {https://doi.org/10.1063/5.0116299} {\bibfield
  {journal} {\bibinfo  {journal} {The Journal of Chemical Physics}\ }\textbf
  {\bibinfo {volume} {158}},\ \bibinfo {pages} {034103} (\bibinfo {year}
  {2023})}\BibitemShut {NoStop}%
\bibitem [{\citenamefont {Zwanzig}(2001)}]{Zwanzig2001}%
  \BibitemOpen
  \bibfield  {author} {\bibinfo {author} {\bibfnamefont {R.}~\bibnamefont
  {Zwanzig}},\ }\href@noop {} {\emph {\bibinfo {title} {Nonequilibrium
  Statistical Mechanics}}}\ (\bibinfo  {publisher} {{Oxford University
  Press}},\ \bibinfo {address} {{New York}},\ \bibinfo {year}
  {2001})\BibitemShut {NoStop}%
\bibitem [{\citenamefont
  {Rudzinski}(2019)}]{rudzinskiRecentProgressChemicallyspecific2019}%
  \BibitemOpen
  \bibfield  {author} {\bibinfo {author} {\bibfnamefont {J.~F.}\ \bibnamefont
  {Rudzinski}},\ }\bibfield  {title} {\enquote {\bibinfo {title} {Recent
  progress towards chemically-specific coarse-grained simulation models with
  consistent dynamical properties},}\ }\href@noop {} {\bibfield  {journal}
  {\bibinfo  {journal} {Computation}\ }\textbf {\bibinfo {volume} {7}},\
  \bibinfo {pages} {42} (\bibinfo {year} {2019})}\BibitemShut {NoStop}%
\bibitem [{\citenamefont
  {Schilling}(2022)}]{schillingCoarsegrainedModellingOut2022}%
  \BibitemOpen
  \bibfield  {author} {\bibinfo {author} {\bibfnamefont {T.}~\bibnamefont
  {Schilling}},\ }\bibfield  {title} {\enquote {\bibinfo {title}
  {Coarse-grained modelling out of equilibrium},}\ }\href@noop {} {\bibfield
  {journal} {\bibinfo  {journal} {Physics Reports}\ }\textbf {\bibinfo {volume}
  {972}},\ \bibinfo {pages} {1--45} (\bibinfo {year} {2022})}\BibitemShut
  {NoStop}%
\bibitem [{\citenamefont {Klippenstein}\ \emph {et~al.}(2021)\citenamefont
  {Klippenstein}, \citenamefont {Tripathy}, \citenamefont {Jung}, \citenamefont
  {Schmid},\ and\ \citenamefont {{van der
  Vegt}}}]{klippensteinIntroducingMemoryCoarseGrained2021}%
  \BibitemOpen
  \bibfield  {author} {\bibinfo {author} {\bibfnamefont {V.}~\bibnamefont
  {Klippenstein}}, \bibinfo {author} {\bibfnamefont {M.}~\bibnamefont
  {Tripathy}}, \bibinfo {author} {\bibfnamefont {G.}~\bibnamefont {Jung}},
  \bibinfo {author} {\bibfnamefont {F.}~\bibnamefont {Schmid}},\ and\ \bibinfo
  {author} {\bibfnamefont {N.~F.~A.}\ \bibnamefont {{van der Vegt}}},\
  }\bibfield  {title} {\enquote {\bibinfo {title} {Introducing {{Memory}} in
  {{Coarse-Grained Molecular Simulations}}},}\ }\href
  {https://doi.org/10.1021/acs.jpcb.1c01120} {\bibfield  {journal} {\bibinfo
  {journal} {The Journal of Physical Chemistry B}\ }\textbf {\bibinfo {volume}
  {125}},\ \bibinfo {pages} {4931--4954} (\bibinfo {year} {2021})}\BibitemShut
  {NoStop}%
\bibitem [{\citenamefont {Tsch{\"o}p}\ \emph {et~al.}(1998)\citenamefont
  {Tsch{\"o}p}, \citenamefont {Kremer}, \citenamefont {Batoulis}, \citenamefont
  {B{\"u}rger},\ and\ \citenamefont {Hahn}}]{tschopSimulationPolymerMelts1998}%
  \BibitemOpen
  \bibfield  {author} {\bibinfo {author} {\bibfnamefont {W.}~\bibnamefont
  {Tsch{\"o}p}}, \bibinfo {author} {\bibfnamefont {K.}~\bibnamefont {Kremer}},
  \bibinfo {author} {\bibfnamefont {J.}~\bibnamefont {Batoulis}}, \bibinfo
  {author} {\bibfnamefont {T.}~\bibnamefont {B{\"u}rger}},\ and\ \bibinfo
  {author} {\bibfnamefont {O.}~\bibnamefont {Hahn}},\ }\bibfield  {title}
  {\enquote {\bibinfo {title} {Simulation of polymer melts. {{I}}.
  {{Coarse-graining}} procedure for polycarbonates},}\ }\href
  {https://doi.org/10.1002/(SICI)1521-4044(199802)49:2/3<61::AID-APOL61>3.0.CO;2-V}
  {\bibfield  {journal} {\bibinfo  {journal} {Acta Polymerica}\ }\textbf
  {\bibinfo {volume} {49}},\ \bibinfo {pages} {61--74} (\bibinfo {year}
  {1998})}\BibitemShut {NoStop}%
\bibitem [{\citenamefont {Lopez}\ \emph {et~al.}(2002)\citenamefont {Lopez},
  \citenamefont {Moore}, \citenamefont {Shelley}, \citenamefont {Shelley},\
  and\ \citenamefont {Klein}}]{lopezComputerSimulationStudies2002}%
  \BibitemOpen
  \bibfield  {author} {\bibinfo {author} {\bibfnamefont {C.~F.}\ \bibnamefont
  {Lopez}}, \bibinfo {author} {\bibfnamefont {P.~B.}\ \bibnamefont {Moore}},
  \bibinfo {author} {\bibfnamefont {J.~C.}\ \bibnamefont {Shelley}}, \bibinfo
  {author} {\bibfnamefont {M.~Y.}\ \bibnamefont {Shelley}},\ and\ \bibinfo
  {author} {\bibfnamefont {M.~L.}\ \bibnamefont {Klein}},\ }\bibfield  {title}
  {\enquote {\bibinfo {title} {Computer simulation studies of biomembranes
  using a coarse grain model},}\ }\href@noop {} {\bibfield  {journal} {\bibinfo
   {journal} {Computer Physics Communications}\ }\textbf {\bibinfo {volume}
  {147}},\ \bibinfo {pages} {1--6} (\bibinfo {year} {2002})}\BibitemShut
  {NoStop}%
\bibitem [{\citenamefont {Depa}\ and\ \citenamefont
  {Maranas}(2005)}]{depaSpeedDynamicObservables2005}%
  \BibitemOpen
  \bibfield  {author} {\bibinfo {author} {\bibfnamefont {P.~K.}\ \bibnamefont
  {Depa}}\ and\ \bibinfo {author} {\bibfnamefont {J.~K.}\ \bibnamefont
  {Maranas}},\ }\bibfield  {title} {\enquote {\bibinfo {title} {Speed up of
  dynamic observables in coarse-grained molecular-dynamics simulations of
  unentangled polymers},}\ }\href {https://doi.org/10.1063/1.1997150}
  {\bibfield  {journal} {\bibinfo  {journal} {The Journal of Chemical Physics}\
  }\textbf {\bibinfo {volume} {123}},\ \bibinfo {pages} {094901} (\bibinfo
  {year} {2005})}\BibitemShut {NoStop}%
\bibitem [{\citenamefont {Fritz}\ \emph {et~al.}(2011)\citenamefont {Fritz},
  \citenamefont {Koschke}, \citenamefont {A.~Harmandaris}, \citenamefont
  {van~der Vegt},\ and\ \citenamefont
  {Kremer}}]{fritzMultiscaleModelingSoft2011}%
  \BibitemOpen
  \bibfield  {author} {\bibinfo {author} {\bibfnamefont {D.}~\bibnamefont
  {Fritz}}, \bibinfo {author} {\bibfnamefont {K.}~\bibnamefont {Koschke}},
  \bibinfo {author} {\bibfnamefont {V.}~\bibnamefont {A.~Harmandaris}},
  \bibinfo {author} {\bibfnamefont {N.~F.~A.}\ \bibnamefont {van~der Vegt}},\
  and\ \bibinfo {author} {\bibfnamefont {K.}~\bibnamefont {Kremer}},\
  }\bibfield  {title} {\enquote {\bibinfo {title} {Multiscale modeling of soft
  matter: Scaling of dynamics},}\ }\href {https://doi.org/10.1039/C1CP20247B}
  {\bibfield  {journal} {\bibinfo  {journal} {Physical Chemistry Chemical
  Physics}\ }\textbf {\bibinfo {volume} {13}},\ \bibinfo {pages} {10412--10420}
  (\bibinfo {year} {2011})}\BibitemShut {NoStop}%
\bibitem [{\citenamefont {Johnson}\ and\ \citenamefont
  {Phelan}(2023)}]{johnsonComparisonFrictionParametrization2023}%
  \BibitemOpen
  \bibfield  {author} {\bibinfo {author} {\bibfnamefont {L.~C.}\ \bibnamefont
  {Johnson}}\ and\ \bibinfo {author} {\bibfnamefont {F.~R.~J.}\ \bibnamefont
  {Phelan}},\ }\bibfield  {title} {\enquote {\bibinfo {title} {Comparison of
  {{Friction Parametrization}} from {{Dynamics}} and {{Material Properties}}
  for a {{Coarse-Grained Polymer Melt}}},}\ }\href
  {https://doi.org/10.1021/acs.jpcb.3c03273} {\bibfield  {journal} {\bibinfo
  {journal} {The Journal of Physical Chemistry B}\ }\textbf {\bibinfo {volume}
  {127}},\ \bibinfo {pages} {7054--7069} (\bibinfo {year} {2023})}\BibitemShut
  {NoStop}%
\bibitem [{\citenamefont {Lei}, \citenamefont {Caswell},\ and\ \citenamefont
  {Karniadakis}(2010)}]{leiDirectConstructionMesoscopic2010}%
  \BibitemOpen
  \bibfield  {author} {\bibinfo {author} {\bibfnamefont {H.}~\bibnamefont
  {Lei}}, \bibinfo {author} {\bibfnamefont {B.}~\bibnamefont {Caswell}},\ and\
  \bibinfo {author} {\bibfnamefont {G.~E.}\ \bibnamefont {Karniadakis}},\
  }\bibfield  {title} {\enquote {\bibinfo {title} {Direct construction of
  mesoscopic models from microscopic simulations},}\ }\href
  {https://doi.org/10.1103/PhysRevE.81.026704} {\bibfield  {journal} {\bibinfo
  {journal} {Physical Review E}\ }\textbf {\bibinfo {volume} {81}},\ \bibinfo
  {pages} {026704} (\bibinfo {year} {2010})}\BibitemShut {NoStop}%
\bibitem [{\citenamefont {Li}\ \emph {et~al.}(2015)\citenamefont {Li},
  \citenamefont {Bian}, \citenamefont {Li},\ and\ \citenamefont
  {Karniadakis}}]{liIncorporationMemoryEffects2015}%
  \BibitemOpen
  \bibfield  {author} {\bibinfo {author} {\bibfnamefont {Z.}~\bibnamefont
  {Li}}, \bibinfo {author} {\bibfnamefont {X.}~\bibnamefont {Bian}}, \bibinfo
  {author} {\bibfnamefont {X.}~\bibnamefont {Li}},\ and\ \bibinfo {author}
  {\bibfnamefont {G.~E.}\ \bibnamefont {Karniadakis}},\ }\bibfield  {title}
  {\enquote {\bibinfo {title} {Incorporation of memory effects in
  coarse-grained modeling via the {{Mori-Zwanzig}} formalism},}\ }\href
  {https://doi.org/10.1063/1.4935490} {\bibfield  {journal} {\bibinfo
  {journal} {The Journal of Chemical Physics}\ }\textbf {\bibinfo {volume}
  {143}},\ \bibinfo {pages} {243128} (\bibinfo {year} {2015})}\BibitemShut
  {NoStop}%
\bibitem [{\citenamefont {Li}\ \emph {et~al.}(2017)\citenamefont {Li},
  \citenamefont {Lee}, \citenamefont {Darve},\ and\ \citenamefont
  {Karniadakis}}]{liComputingNonMarkovianCoarsegrained2017a}%
  \BibitemOpen
  \bibfield  {author} {\bibinfo {author} {\bibfnamefont {Z.}~\bibnamefont
  {Li}}, \bibinfo {author} {\bibfnamefont {H.~S.}\ \bibnamefont {Lee}},
  \bibinfo {author} {\bibfnamefont {E.}~\bibnamefont {Darve}},\ and\ \bibinfo
  {author} {\bibfnamefont {G.~E.}\ \bibnamefont {Karniadakis}},\ }\bibfield
  {title} {\enquote {\bibinfo {title} {Computing the non-{{Markovian}}
  coarse-grained interactions derived from the {{Mori}}{\textendash}{{Zwanzig}}
  formalism in molecular systems: {{Application}} to polymer melts},}\ }\href
  {https://doi.org/10.1063/1.4973347} {\bibfield  {journal} {\bibinfo
  {journal} {The Journal of Chemical Physics}\ }\textbf {\bibinfo {volume}
  {146}},\ \bibinfo {pages} {014104} (\bibinfo {year} {2017})}\BibitemShut
  {NoStop}%
\bibitem [{\citenamefont {Han}, \citenamefont {Dama},\ and\ \citenamefont
  {Voth}(2018)}]{hanMesoscopicCoarsegrainedRepresentations2018}%
  \BibitemOpen
  \bibfield  {author} {\bibinfo {author} {\bibfnamefont {Y.}~\bibnamefont
  {Han}}, \bibinfo {author} {\bibfnamefont {J.~F.}\ \bibnamefont {Dama}},\ and\
  \bibinfo {author} {\bibfnamefont {G.~A.}\ \bibnamefont {Voth}},\ }\bibfield
  {title} {\enquote {\bibinfo {title} {Mesoscopic coarse-grained
  representations of fluids rigorously derived from atomistic models},}\ }\href
  {https://doi.org/10.1063/1.5039738} {\bibfield  {journal} {\bibinfo
  {journal} {Journal of Chemical Physics}\ }\textbf {\bibinfo {volume} {149}},\
  \bibinfo {pages} {44104} (\bibinfo {year} {2018})}\BibitemShut {NoStop}%
\bibitem [{\citenamefont {Han}, \citenamefont {Jin},\ and\ \citenamefont
  {Voth}(2021)}]{hanConstructingManybodyDissipative2021}%
  \BibitemOpen
  \bibfield  {author} {\bibinfo {author} {\bibfnamefont {Y.}~\bibnamefont
  {Han}}, \bibinfo {author} {\bibfnamefont {J.}~\bibnamefont {Jin}},\ and\
  \bibinfo {author} {\bibfnamefont {G.~A.}\ \bibnamefont {Voth}},\ }\bibfield
  {title} {\enquote {\bibinfo {title} {Constructing many-body dissipative
  particle dynamics models of fluids from bottom-up coarse-graining},}\ }\href
  {https://doi.org/10.1063/5.0035184} {\bibfield  {journal} {\bibinfo
  {journal} {The Journal of Chemical Physics}\ }\textbf {\bibinfo {volume}
  {154}},\ \bibinfo {pages} {084122} (\bibinfo {year} {2021})}\BibitemShut
  {NoStop}%
\bibitem [{\citenamefont {Klippenstein}\ and\ \citenamefont {{van der
  Vegt}}(2021)}]{klippensteinCrosscorrelationCorrectedFriction2021}%
  \BibitemOpen
  \bibfield  {author} {\bibinfo {author} {\bibfnamefont {V.}~\bibnamefont
  {Klippenstein}}\ and\ \bibinfo {author} {\bibfnamefont {N.~F.~A.}\
  \bibnamefont {{van der Vegt}}},\ }\bibfield  {title} {\enquote {\bibinfo
  {title} {Cross-correlation corrected friction in (generalized) {{Langevin}}
  models},}\ }\href {https://doi.org/10.1063/5.0049324} {\bibfield  {journal}
  {\bibinfo  {journal} {The Journal of Chemical Physics}\ }\textbf {\bibinfo
  {volume} {154}},\ \bibinfo {pages} {191102} (\bibinfo {year}
  {2021})}\BibitemShut {NoStop}%
\bibitem [{\citenamefont {Klippenstein}\ and\ \citenamefont {{van der
  Vegt}}(2022)}]{klippensteinCrosscorrelationCorrectedFriction2022}%
  \BibitemOpen
  \bibfield  {author} {\bibinfo {author} {\bibfnamefont {V.}~\bibnamefont
  {Klippenstein}}\ and\ \bibinfo {author} {\bibfnamefont {N.~F.~A.}\
  \bibnamefont {{van der Vegt}}},\ }\bibfield  {title} {\enquote {\bibinfo
  {title} {Cross-correlation corrected friction in generalized {{Langevin}}
  models: {{Application}} to the continuous {{Asakura}}{\textendash}{{Oosawa}}
  model},}\ }\href {https://doi.org/10.1063/5.0093056} {\bibfield  {journal}
  {\bibinfo  {journal} {The Journal of Chemical Physics}\ }\textbf {\bibinfo
  {volume} {157}},\ \bibinfo {pages} {044103} (\bibinfo {year}
  {2022})}\BibitemShut {NoStop}%
\bibitem [{\citenamefont {Klippenstein}\ and\ \citenamefont {{van der
  Vegt}}(2023)}]{klippensteinBottomUpInformedIteratively2023}%
  \BibitemOpen
  \bibfield  {author} {\bibinfo {author} {\bibfnamefont {V.}~\bibnamefont
  {Klippenstein}}\ and\ \bibinfo {author} {\bibfnamefont {N.~F.~A.}\
  \bibnamefont {{van der Vegt}}},\ }\bibfield  {title} {\enquote {\bibinfo
  {title} {Bottom-{{Up Informed}} and {{Iteratively Optimized Coarse-Grained
  Non-Markovian Water Models}} with {{Accurate Dynamics}}},}\ }\href
  {https://doi.org/10.1021/acs.jctc.2c00871} {\bibfield  {journal} {\bibinfo
  {journal} {Journal of Chemical Theory and Computation}\ }\textbf {\bibinfo
  {volume} {19}},\ \bibinfo {pages} {1099--1110} (\bibinfo {year}
  {2023})}\BibitemShut {NoStop}%
\bibitem [{\citenamefont {Tripathy}, \citenamefont {Klippenstein},\ and\
  \citenamefont {{van der
  Vegt}}(2023)}]{tripathyDynamicalCoarsegrainedModels2023}%
  \BibitemOpen
  \bibfield  {author} {\bibinfo {author} {\bibfnamefont {M.}~\bibnamefont
  {Tripathy}}, \bibinfo {author} {\bibfnamefont {V.}~\bibnamefont
  {Klippenstein}},\ and\ \bibinfo {author} {\bibfnamefont {N.~F.~A.}\
  \bibnamefont {{van der Vegt}}},\ }\bibfield  {title} {\enquote {\bibinfo
  {title} {Dynamical coarse-grained models of molecular liquids and their ideal
  and non-ideal mixtures},}\ }\href {https://doi.org/10.1063/5.0163097}
  {\bibfield  {journal} {\bibinfo  {journal} {The Journal of Chemical Physics}\
  }\textbf {\bibinfo {volume} {159}},\ \bibinfo {pages} {094904} (\bibinfo
  {year} {2023})}\BibitemShut {NoStop}%
\bibitem [{\citenamefont {Mori}(1965)}]{Mori1965a}%
  \BibitemOpen
  \bibfield  {author} {\bibinfo {author} {\bibfnamefont {H.}~\bibnamefont
  {Mori}},\ }\bibfield  {title} {\enquote {\bibinfo {title} {Transport,
  collective motion, and brownian motion*)},}\ }\href@noop {} {\bibfield
  {journal} {\bibinfo  {journal} {Progress of Theoretical Physics}\ }\textbf
  {\bibinfo {volume} {33}},\ \bibinfo {pages} {423} (\bibinfo {year}
  {1965})}\BibitemShut {NoStop}%
\bibitem [{\citenamefont {Izvekov}\ and\ \citenamefont
  {Voth}(2006)}]{izvekovModelingRealDynamics2006}%
  \BibitemOpen
  \bibfield  {author} {\bibinfo {author} {\bibfnamefont {S.}~\bibnamefont
  {Izvekov}}\ and\ \bibinfo {author} {\bibfnamefont {G.~A.}\ \bibnamefont
  {Voth}},\ }\bibfield  {title} {\enquote {\bibinfo {title} {Modeling real
  dynamics in the coarse-grained representation of condensed phase systems},}\
  }\href {https://doi.org/10.1063/1.2360580} {\bibfield  {journal} {\bibinfo
  {journal} {The Journal of Chemical Physics}\ }\textbf {\bibinfo {volume}
  {125}},\ \bibinfo {pages} {151101} (\bibinfo {year} {2006})}\BibitemShut
  {NoStop}%
\bibitem [{\citenamefont {Hij{\'o}n}\ \emph {et~al.}(2010)\citenamefont
  {Hij{\'o}n}, \citenamefont {Espa{\~n}ol}, \citenamefont {{Vanden-Eijnden}},\
  and\ \citenamefont {{Delgado-Buscalioni}}}]{hijonMoriZwanzigFormalism2010}%
  \BibitemOpen
  \bibfield  {author} {\bibinfo {author} {\bibfnamefont {C.}~\bibnamefont
  {Hij{\'o}n}}, \bibinfo {author} {\bibfnamefont {P.}~\bibnamefont
  {Espa{\~n}ol}}, \bibinfo {author} {\bibfnamefont {E.}~\bibnamefont
  {{Vanden-Eijnden}}},\ and\ \bibinfo {author} {\bibfnamefont {R.}~\bibnamefont
  {{Delgado-Buscalioni}}},\ }\bibfield  {title} {\enquote {\bibinfo {title}
  {Mori{\textendash}{{Zwanzig}} formalism as a practical computational tool},}\
  }\href {https://doi.org/10.1039/B902479B} {\bibfield  {journal} {\bibinfo
  {journal} {Faraday Discussions}\ }\textbf {\bibinfo {volume} {144}},\
  \bibinfo {pages} {301--322} (\bibinfo {year} {2010})}\BibitemShut {NoStop}%
\bibitem [{\citenamefont {Jung}, \citenamefont {Hanke},\ and\ \citenamefont
  {Schmid}(2018)}]{jungGeneralizedLangevinDynamics2018}%
  \BibitemOpen
  \bibfield  {author} {\bibinfo {author} {\bibfnamefont {G.}~\bibnamefont
  {Jung}}, \bibinfo {author} {\bibfnamefont {M.}~\bibnamefont {Hanke}},\ and\
  \bibinfo {author} {\bibfnamefont {F.}~\bibnamefont {Schmid}},\ }\bibfield
  {title} {\enquote {\bibinfo {title} {Generalized {{Langevin}} dynamics:
  Construction and numerical integration of non-{{Markovian}} particle-based
  models},}\ }\href {https://doi.org/10.1039/C8SM01817K} {\bibfield  {journal}
  {\bibinfo  {journal} {Soft Matter}\ }\textbf {\bibinfo {volume} {14}},\
  \bibinfo {pages} {9368--9382} (\bibinfo {year} {2018})}\BibitemShut {NoStop}%
\bibitem [{\citenamefont {Wang}, \citenamefont {Ma},\ and\ \citenamefont
  {Pan}(2020)}]{wangDatadrivenCoarsegrainedModeling2020}%
  \BibitemOpen
  \bibfield  {author} {\bibinfo {author} {\bibfnamefont {S.}~\bibnamefont
  {Wang}}, \bibinfo {author} {\bibfnamefont {Z.}~\bibnamefont {Ma}},\ and\
  \bibinfo {author} {\bibfnamefont {W.}~\bibnamefont {Pan}},\ }\bibfield
  {title} {\enquote {\bibinfo {title} {Data-driven coarse-grained modeling of
  polymers in solution with structural and dynamic properties conserved},}\
  }\href {https://doi.org/10.1039/D0SM01019G} {\bibfield  {journal} {\bibinfo
  {journal} {Soft Matter}\ }\textbf {\bibinfo {volume} {16}},\ \bibinfo {pages}
  {8330--8344} (\bibinfo {year} {2020})}\BibitemShut {NoStop}%
\bibitem [{\citenamefont {Jung}, \citenamefont {Hanke},\ and\ \citenamefont
  {Schmid}(2017)}]{jungIterativeReconstructionMemory2017}%
  \BibitemOpen
  \bibfield  {author} {\bibinfo {author} {\bibfnamefont {G.}~\bibnamefont
  {Jung}}, \bibinfo {author} {\bibfnamefont {M.}~\bibnamefont {Hanke}},\ and\
  \bibinfo {author} {\bibfnamefont {F.}~\bibnamefont {Schmid}},\ }\bibfield
  {title} {\enquote {\bibinfo {title} {Iterative reconstruction of memory
  kernels},}\ }\href {https://doi.org/10.1021/acs.jctc.7b00274} {\bibfield
  {journal} {\bibinfo  {journal} {Journal of Chemical Theory and Computation}\
  }\textbf {\bibinfo {volume} {13}},\ \bibinfo {pages} {2481--2488} (\bibinfo
  {year} {2017})}\BibitemShut {NoStop}%
\bibitem [{\citenamefont {Meyer}, \citenamefont {Pelagejcev},\ and\
  \citenamefont
  {Schilling}(2020)}]{meyerNonMarkovianOutofequilibriumDynamics2020}%
  \BibitemOpen
  \bibfield  {author} {\bibinfo {author} {\bibfnamefont {H.}~\bibnamefont
  {Meyer}}, \bibinfo {author} {\bibfnamefont {P.}~\bibnamefont {Pelagejcev}},\
  and\ \bibinfo {author} {\bibfnamefont {T.}~\bibnamefont {Schilling}},\
  }\bibfield  {title} {\enquote {\bibinfo {title} {Non-{{Markovian}}
  out-of-equilibrium dynamics: {{A}} general numerical procedure to construct
  time-dependent memory kernels for coarse-grained observables},}\ }\href
  {https://doi.org/10.1209/0295-5075/128/40001} {\bibfield  {journal} {\bibinfo
   {journal} {Europhysics Letters}\ }\textbf {\bibinfo {volume} {128}},\
  \bibinfo {pages} {40001} (\bibinfo {year} {2020})}\BibitemShut {NoStop}%
\bibitem [{\citenamefont
  {Hanke}(2021)}]{hankeMathematicalAnalysisIterative2021}%
  \BibitemOpen
  \bibfield  {author} {\bibinfo {author} {\bibfnamefont {M.}~\bibnamefont
  {Hanke}},\ }\bibfield  {title} {\enquote {\bibinfo {title} {Mathematical
  analysis of some iterative methods for the reconstruction of memory
  kernels},}\ }\href {https://doi.org/10.1553/etna_vol54s483} {\bibfield
  {journal} {\bibinfo  {journal} {Electronic Transactions on Numerical
  Analysis}\ }\textbf {\bibinfo {volume} {54}},\ \bibinfo {pages} {483--498}
  (\bibinfo {year} {2021})}\BibitemShut {NoStop}%
\bibitem [{\citenamefont {Gao}\ and\ \citenamefont
  {Fang}(2011)}]{gaoSemibottomupCoarseGraining2011}%
  \BibitemOpen
  \bibfield  {author} {\bibinfo {author} {\bibfnamefont {L.}~\bibnamefont
  {Gao}}\ and\ \bibinfo {author} {\bibfnamefont {W.}~\bibnamefont {Fang}},\
  }\bibfield  {title} {\enquote {\bibinfo {title} {Semi-bottom-up coarse
  graining of water based on microscopic simulations},}\ }\href
  {https://doi.org/10.1063/1.3658500} {\bibfield  {journal} {\bibinfo
  {journal} {The Journal of Chemical Physics}\ }\textbf {\bibinfo {volume}
  {135}},\ \bibinfo {pages} {184101} (\bibinfo {year} {2011})}\BibitemShut
  {NoStop}%
\bibitem [{\citenamefont {Ceriotti}, \citenamefont {Bussi},\ and\ \citenamefont
  {Parrinello}(2009)}]{ceriottiLangevinEquationColored2009a}%
  \BibitemOpen
  \bibfield  {author} {\bibinfo {author} {\bibfnamefont {M.}~\bibnamefont
  {Ceriotti}}, \bibinfo {author} {\bibfnamefont {G.}~\bibnamefont {Bussi}},\
  and\ \bibinfo {author} {\bibfnamefont {M.}~\bibnamefont {Parrinello}},\
  }\bibfield  {title} {\enquote {\bibinfo {title} {Langevin {{Equation}} with
  {{Colored Noise}} for {{Constant-Temperature Molecular Dynamics
  Simulations}}},}\ }\href {https://doi.org/10.1103/PhysRevLett.102.020601}
  {\bibfield  {journal} {\bibinfo  {journal} {Physical Review Letters}\
  }\textbf {\bibinfo {volume} {102}},\ \bibinfo {pages} {020601} (\bibinfo
  {year} {2009})}\BibitemShut {NoStop}%
\bibitem [{\citenamefont {Ferrario}\ and\ \citenamefont
  {Grigolini}(1979)}]{ferrarioNonMarkovianRelaxationProcess1979}%
  \BibitemOpen
  \bibfield  {author} {\bibinfo {author} {\bibfnamefont {M.}~\bibnamefont
  {Ferrario}}\ and\ \bibinfo {author} {\bibfnamefont {P.}~\bibnamefont
  {Grigolini}},\ }\bibfield  {title} {\enquote {\bibinfo {title} {The
  non-{{Markovian}} relaxation process as a ``contraction'' of a
  multidimensional one of {{Markovian}} type},}\ }\href
  {https://doi.org/10.1063/1.524019} {\bibfield  {journal} {\bibinfo  {journal}
  {Journal of Mathematical Physics}\ }\textbf {\bibinfo {volume} {20}},\
  \bibinfo {pages} {2567--2572} (\bibinfo {year} {1979})}\BibitemShut {NoStop}%
\bibitem [{\citenamefont {Marchesoni}\ and\ \citenamefont
  {Grigolini}(1983)}]{marchesoniExtensionKramersTheory1983}%
  \BibitemOpen
  \bibfield  {author} {\bibinfo {author} {\bibfnamefont {F.}~\bibnamefont
  {Marchesoni}}\ and\ \bibinfo {author} {\bibfnamefont {P.}~\bibnamefont
  {Grigolini}},\ }\bibfield  {title} {\enquote {\bibinfo {title} {On the
  extension of the {{Kramers}} theory of chemical relaxation to the case of
  nonwhite noise},}\ }\href {https://doi.org/10.1063/1.444554} {\bibfield
  {journal} {\bibinfo  {journal} {The Journal of Chemical Physics}\ }\textbf
  {\bibinfo {volume} {78}},\ \bibinfo {pages} {6287--6298} (\bibinfo {year}
  {1983})}\BibitemShut {NoStop}%
\bibitem [{\citenamefont {Shin}\ \emph {et~al.}(2010)\citenamefont {Shin},
  \citenamefont {Kim}, \citenamefont {Talkner},\ and\ \citenamefont
  {Lee}}]{shinBrownianMotionMolecular2010}%
  \BibitemOpen
  \bibfield  {author} {\bibinfo {author} {\bibfnamefont {H.~K.}\ \bibnamefont
  {Shin}}, \bibinfo {author} {\bibfnamefont {C.}~\bibnamefont {Kim}}, \bibinfo
  {author} {\bibfnamefont {P.}~\bibnamefont {Talkner}},\ and\ \bibinfo {author}
  {\bibfnamefont {E.~K.}\ \bibnamefont {Lee}},\ }\bibfield  {title} {\enquote
  {\bibinfo {title} {Brownian motion from molecular dynamics},}\ }\href@noop {}
  {\bibfield  {journal} {\bibinfo  {journal} {Chemical Physics}\ }\textbf
  {\bibinfo {volume} {375}},\ \bibinfo {pages} {316--326} (\bibinfo {year}
  {2010})}\BibitemShut {NoStop}%
\bibitem [{\citenamefont {Ceriotti}, \citenamefont {Bussi},\ and\ \citenamefont
  {Parrinello}(2010)}]{ceriottiColoredNoiseThermostatsCarte2010a}%
  \BibitemOpen
  \bibfield  {author} {\bibinfo {author} {\bibfnamefont {M.}~\bibnamefont
  {Ceriotti}}, \bibinfo {author} {\bibfnamefont {G.}~\bibnamefont {Bussi}},\
  and\ \bibinfo {author} {\bibfnamefont {M.}~\bibnamefont {Parrinello}},\
  }\bibfield  {title} {\enquote {\bibinfo {title} {Colored-{{Noise
  Thermostats}} {\`a} la {{Carte}}},}\ }\href
  {https://doi.org/10.1021/ct900563s} {\bibfield  {journal} {\bibinfo
  {journal} {Journal of Chemical Theory and Computation}\ }\textbf {\bibinfo
  {volume} {6}},\ \bibinfo {pages} {1170--1180} (\bibinfo {year}
  {2010})}\BibitemShut {NoStop}%
\bibitem [{\citenamefont {Wang}, \citenamefont {Li},\ and\ \citenamefont
  {Pan}(2019)}]{wangImplicitsolventCoarsegrainedModeling2019}%
  \BibitemOpen
  \bibfield  {author} {\bibinfo {author} {\bibfnamefont {S.}~\bibnamefont
  {Wang}}, \bibinfo {author} {\bibfnamefont {Z.}~\bibnamefont {Li}},\ and\
  \bibinfo {author} {\bibfnamefont {W.}~\bibnamefont {Pan}},\ }\bibfield
  {title} {\enquote {\bibinfo {title} {Implicit-solvent coarse-grained modeling
  for polymer solutions via {{Mori-Zwanzig}} formalism},}\ }\href
  {https://doi.org/10.1039/c9sm01211g} {\bibfield  {journal} {\bibinfo
  {journal} {Soft matter}\ }\textbf {\bibinfo {volume} {15}},\ \bibinfo {pages}
  {7567--7582} (\bibinfo {year} {2019})}\BibitemShut {NoStop}%
\bibitem [{\citenamefont
  {Vroylandt}(2022)}]{vroylandtDerivationGeneralizedLangevin2022}%
  \BibitemOpen
  \bibfield  {author} {\bibinfo {author} {\bibfnamefont {H.}~\bibnamefont
  {Vroylandt}},\ }\bibfield  {title} {\enquote {\bibinfo {title} {On the
  derivation of the generalized {{Langevin}} equation and the
  fluctuation-dissipation theorem},}\ }\href
  {https://doi.org/10.1209/0295-5075/acab7d} {\bibfield  {journal} {\bibinfo
  {journal} {Europhysics Letters}\ }\textbf {\bibinfo {volume} {140}},\
  \bibinfo {pages} {62003} (\bibinfo {year} {2022})}\BibitemShut {NoStop}%
\bibitem [{\citenamefont {Glatzel}\ and\ \citenamefont
  {Schilling}(2022)}]{glatzelInterplayMemoryPotentials2022}%
  \BibitemOpen
  \bibfield  {author} {\bibinfo {author} {\bibfnamefont {F.}~\bibnamefont
  {Glatzel}}\ and\ \bibinfo {author} {\bibfnamefont {T.}~\bibnamefont
  {Schilling}},\ }\bibfield  {title} {\enquote {\bibinfo {title} {The interplay
  between memory and potentials of mean force: {{A}} discussion on the
  structure of equations of motion for coarse-grained observables},}\ }\href
  {https://doi.org/10.1209/0295-5075/ac35ba} {\bibfield  {journal} {\bibinfo
  {journal} {Europhysics Letters}\ }\textbf {\bibinfo {volume} {136}},\
  \bibinfo {pages} {36001} (\bibinfo {year} {2022})}\BibitemShut {NoStop}%
\bibitem [{\citenamefont {Kowalik}\ \emph {et~al.}(2019)\citenamefont
  {Kowalik}, \citenamefont {Daldrop}, \citenamefont {Kappler}, \citenamefont
  {Schulz}, \citenamefont {Schlaich},\ and\ \citenamefont
  {Netz}}]{kowalikMemorykernelExtractionDifferent2019}%
  \BibitemOpen
  \bibfield  {author} {\bibinfo {author} {\bibfnamefont {B.}~\bibnamefont
  {Kowalik}}, \bibinfo {author} {\bibfnamefont {J.~O.}\ \bibnamefont
  {Daldrop}}, \bibinfo {author} {\bibfnamefont {J.}~\bibnamefont {Kappler}},
  \bibinfo {author} {\bibfnamefont {J.~C.}\ \bibnamefont {Schulz}}, \bibinfo
  {author} {\bibfnamefont {A.}~\bibnamefont {Schlaich}},\ and\ \bibinfo
  {author} {\bibfnamefont {R.~R.}\ \bibnamefont {Netz}},\ }\bibfield  {title}
  {\enquote {\bibinfo {title} {Memory-kernel extraction for different molecular
  solutes in solvents of varying viscosity in confinement},}\ }\href
  {https://doi.org/10.1103/PhysRevE.100.012126} {\bibfield  {journal} {\bibinfo
   {journal} {Physical Review E}\ }\textbf {\bibinfo {volume} {100}},\ \bibinfo
  {pages} {12126} (\bibinfo {year} {2019})}\BibitemShut {NoStop}%
\bibitem [{\citenamefont {Straube}\ \emph {et~al.}(2020)\citenamefont
  {Straube}, \citenamefont {Kowalik}, \citenamefont {Netz},\ and\ \citenamefont
  {H{\"o}fling}}]{straubeRapidOnsetMolecular2020b}%
  \BibitemOpen
  \bibfield  {author} {\bibinfo {author} {\bibfnamefont {A.~V.}\ \bibnamefont
  {Straube}}, \bibinfo {author} {\bibfnamefont {B.~G.}\ \bibnamefont
  {Kowalik}}, \bibinfo {author} {\bibfnamefont {R.~R.}\ \bibnamefont {Netz}},\
  and\ \bibinfo {author} {\bibfnamefont {F.}~\bibnamefont {H{\"o}fling}},\
  }\bibfield  {title} {\enquote {\bibinfo {title} {Rapid onset of molecular
  friction in liquids bridging between the atomistic and hydrodynamic
  pictures},}\ }\href {https://doi.org/10.1038/s42005-020-0389-0} {\bibfield
  {journal} {\bibinfo  {journal} {Communications Physics}\ }\textbf {\bibinfo
  {volume} {3}},\ \bibinfo {pages} {1--11} (\bibinfo {year}
  {2020})}\BibitemShut {NoStop}%
\bibitem [{\citenamefont {Reinhardt}, \citenamefont {Hoffmann},\ and\
  \citenamefont {Gerlach}(2013)}]{reinhardtNichtlineareOptimierung2013}%
  \BibitemOpen
  \bibfield  {author} {\bibinfo {author} {\bibfnamefont {R.}~\bibnamefont
  {Reinhardt}}, \bibinfo {author} {\bibfnamefont {A.}~\bibnamefont
  {Hoffmann}},\ and\ \bibinfo {author} {\bibfnamefont {T.}~\bibnamefont
  {Gerlach}},\ }\href {https://doi.org/10.1007/978-3-8274-2949-0} {\emph
  {\bibinfo {title} {{Nichtlineare Optimierung}}}}\ (\bibinfo  {publisher}
  {{Springer}},\ \bibinfo {address} {{Berlin, Heidelberg}},\ \bibinfo {year}
  {2013})\BibitemShut {NoStop}%
\bibitem [{\citenamefont {Engl}, \citenamefont {Hanke},\ and\ \citenamefont
  {Neubauer}(1996)}]{englRegularizationInverseProblems1996}%
  \BibitemOpen
  \bibfield  {author} {\bibinfo {author} {\bibfnamefont {H.~W.}\ \bibnamefont
  {Engl}}, \bibinfo {author} {\bibfnamefont {M.}~\bibnamefont {Hanke}},\ and\
  \bibinfo {author} {\bibfnamefont {A.}~\bibnamefont {Neubauer}},\ }\href@noop
  {} {\emph {\bibinfo {title} {Regularization of {{Inverse Problems}}}}}\
  (\bibinfo  {publisher} {{Springer Science \& Business Media}},\ \bibinfo
  {year} {1996})\BibitemShut {NoStop}%
\bibitem [{\citenamefont {Deichmann}\ and\ \citenamefont {{van der
  Vegt}}(2018)}]{deichmannBottomupApproachRepresent2018}%
  \BibitemOpen
  \bibfield  {author} {\bibinfo {author} {\bibfnamefont {G.}~\bibnamefont
  {Deichmann}}\ and\ \bibinfo {author} {\bibfnamefont {N.~F.~A.}\ \bibnamefont
  {{van der Vegt}}},\ }\bibfield  {title} {\enquote {\bibinfo {title}
  {Bottom-up approach to represent dynamic properties in coarse-grained
  molecular simulations},}\ }\href {https://doi.org/10.1063/1.5064369}
  {\bibfield  {journal} {\bibinfo  {journal} {The Journal of Chemical Physics}\
  }\textbf {\bibinfo {volume} {149}},\ \bibinfo {pages} {244114} (\bibinfo
  {year} {2018})}\BibitemShut {NoStop}%
\bibitem [{\citenamefont {Tr{\'e}ment}\ \emph {et~al.}(2014)\citenamefont
  {Tr{\'e}ment}, \citenamefont {Schnell}, \citenamefont {Petitjean},
  \citenamefont {Couty},\ and\ \citenamefont
  {Rousseau}}]{trementConservativeDissipativeForce2014}%
  \BibitemOpen
  \bibfield  {author} {\bibinfo {author} {\bibfnamefont {S.}~\bibnamefont
  {Tr{\'e}ment}}, \bibinfo {author} {\bibfnamefont {B.}~\bibnamefont
  {Schnell}}, \bibinfo {author} {\bibfnamefont {L.}~\bibnamefont {Petitjean}},
  \bibinfo {author} {\bibfnamefont {M.}~\bibnamefont {Couty}},\ and\ \bibinfo
  {author} {\bibfnamefont {B.}~\bibnamefont {Rousseau}},\ }\bibfield  {title}
  {\enquote {\bibinfo {title} {Conservative and dissipative force field for
  simulation of coarse-grained alkane molecules: {{A}} bottom-up approach},}\
  }\href {https://doi.org/10.1063/1.4870394} {\bibfield  {journal} {\bibinfo
  {journal} {The Journal of Chemical Physics}\ }\textbf {\bibinfo {volume}
  {140}},\ \bibinfo {pages} {134113} (\bibinfo {year} {2014})}\BibitemShut
  {NoStop}%
\bibitem [{\citenamefont {Deichmann}, \citenamefont {Marcon},\ and\
  \citenamefont {{van der
  Vegt}}(2014)}]{deichmannBottomupDerivationConservative2014}%
  \BibitemOpen
  \bibfield  {author} {\bibinfo {author} {\bibfnamefont {G.}~\bibnamefont
  {Deichmann}}, \bibinfo {author} {\bibfnamefont {V.}~\bibnamefont {Marcon}},\
  and\ \bibinfo {author} {\bibfnamefont {N.~F.~A.}\ \bibnamefont {{van der
  Vegt}}},\ }\bibfield  {title} {\enquote {\bibinfo {title} {Bottom-up
  derivation of conservative and dissipative interactions for coarse-grained
  molecular liquids with the conditional reversible work method},}\ }\href
  {https://doi.org/10.1063/1.4903454} {\bibfield  {journal} {\bibinfo
  {journal} {The Journal of Chemical Physics}\ }\textbf {\bibinfo {volume}
  {141}},\ \bibinfo {pages} {224109} (\bibinfo {year} {2014})}\BibitemShut
  {NoStop}%
\bibitem [{\citenamefont {Lemarchand}, \citenamefont {Couty},\ and\
  \citenamefont {Rousseau}(2017)}]{lemarchandCoarsegrainedSimulationsCis2017}%
  \BibitemOpen
  \bibfield  {author} {\bibinfo {author} {\bibfnamefont {C.~A.}\ \bibnamefont
  {Lemarchand}}, \bibinfo {author} {\bibfnamefont {M.}~\bibnamefont {Couty}},\
  and\ \bibinfo {author} {\bibfnamefont {B.}~\bibnamefont {Rousseau}},\
  }\bibfield  {title} {\enquote {\bibinfo {title} {Coarse-grained simulations
  of cis- and trans-polybutadiene: {{A}} bottom-up approach},}\ }\href
  {https://doi.org/10.1063/1.4975652} {\bibfield  {journal} {\bibinfo
  {journal} {The Journal of Chemical Physics}\ }\textbf {\bibinfo {volume}
  {146}},\ \bibinfo {pages} {074904} (\bibinfo {year} {2017})}\BibitemShut
  {NoStop}%
\bibitem [{\citenamefont {Hoogerbrugge}\ and\ \citenamefont
  {Koelman}(1992)}]{hoogerbruggeSimulatingMicroscopicHydrodynamic1992}%
  \BibitemOpen
  \bibfield  {author} {\bibinfo {author} {\bibfnamefont {P.~J.}\ \bibnamefont
  {Hoogerbrugge}}\ and\ \bibinfo {author} {\bibfnamefont {J.~M. V.~A.}\
  \bibnamefont {Koelman}},\ }\bibfield  {title} {\enquote {\bibinfo {title}
  {Simulating {{Microscopic Hydrodynamic Phenomena}} with {{Dissipative
  Particle Dynamics}}},}\ }\href {https://doi.org/10.1209/0295-5075/19/3/001}
  {\bibfield  {journal} {\bibinfo  {journal} {Europhysics Letters}\ }\textbf
  {\bibinfo {volume} {19}},\ \bibinfo {pages} {155} (\bibinfo {year}
  {1992})}\BibitemShut {NoStop}%
\bibitem [{\citenamefont {Groot}\ and\ \citenamefont
  {Warren}(1997)}]{grootDissipativeParticleDynamics1997}%
  \BibitemOpen
  \bibfield  {author} {\bibinfo {author} {\bibfnamefont {R.~D.}\ \bibnamefont
  {Groot}}\ and\ \bibinfo {author} {\bibfnamefont {P.~B.}\ \bibnamefont
  {Warren}},\ }\bibfield  {title} {\enquote {\bibinfo {title} {Dissipative
  particle dynamics: {{Bridging}} the gap between atomistic and mesoscopic
  simulation},}\ }\href {https://doi.org/10.1063/1.474784} {\bibfield
  {journal} {\bibinfo  {journal} {The Journal of Chemical Physics}\ }\textbf
  {\bibinfo {volume} {107}},\ \bibinfo {pages} {4423--4435} (\bibinfo {year}
  {1997})}\BibitemShut {NoStop}%
\bibitem [{\citenamefont {Eriksson}\ \emph {et~al.}(2008)\citenamefont
  {Eriksson}, \citenamefont {Jacobi}, \citenamefont {Nystr{\"o}m},\ and\
  \citenamefont {Tunstr{\o}m}}]{erikssonEffectiveThermostatInduced2008}%
  \BibitemOpen
  \bibfield  {author} {\bibinfo {author} {\bibfnamefont {A.}~\bibnamefont
  {Eriksson}}, \bibinfo {author} {\bibfnamefont {M.~N.}\ \bibnamefont
  {Jacobi}}, \bibinfo {author} {\bibfnamefont {J.}~\bibnamefont
  {Nystr{\"o}m}},\ and\ \bibinfo {author} {\bibfnamefont {K.}~\bibnamefont
  {Tunstr{\o}m}},\ }\bibfield  {title} {\enquote {\bibinfo {title} {Effective
  thermostat induced by coarse graining of simple point charge water},}\ }\href
  {https://doi.org/10.1063/1.2953320} {\bibfield  {journal} {\bibinfo
  {journal} {The Journal of Chemical Physics}\ }\textbf {\bibinfo {volume}
  {129}},\ \bibinfo {pages} {024106} (\bibinfo {year} {2008})}\BibitemShut
  {NoStop}%
\bibitem [{\citenamefont {Lei}, \citenamefont {Baker},\ and\ \citenamefont
  {Li}(2016)}]{leiDatadrivenParameterizationGeneralized2016}%
  \BibitemOpen
  \bibfield  {author} {\bibinfo {author} {\bibfnamefont {H.}~\bibnamefont
  {Lei}}, \bibinfo {author} {\bibfnamefont {N.~A.}\ \bibnamefont {Baker}},\
  and\ \bibinfo {author} {\bibfnamefont {X.}~\bibnamefont {Li}},\ }\bibfield
  {title} {\enquote {\bibinfo {title} {Data-driven parameterization of the
  generalized langevin equation},}\ }\href@noop {} {\bibfield  {journal}
  {\bibinfo  {journal} {Proceedings of the National Academy of Sciences}\
  }\textbf {\bibinfo {volume} {113}},\ \bibinfo {pages} {14183--14188}
  (\bibinfo {year} {2016})}\BibitemShut {NoStop}%
\bibitem [{\citenamefont {Bockius}\ \emph {et~al.}(2021)\citenamefont
  {Bockius}, \citenamefont {Shea}, \citenamefont {Jung}, \citenamefont
  {Schmid},\ and\ \citenamefont {Hanke}}]{bockiusModelReductionTechniques2021}%
  \BibitemOpen
  \bibfield  {author} {\bibinfo {author} {\bibfnamefont {N.}~\bibnamefont
  {Bockius}}, \bibinfo {author} {\bibfnamefont {J.}~\bibnamefont {Shea}},
  \bibinfo {author} {\bibfnamefont {G.}~\bibnamefont {Jung}}, \bibinfo {author}
  {\bibfnamefont {F.}~\bibnamefont {Schmid}},\ and\ \bibinfo {author}
  {\bibfnamefont {M.}~\bibnamefont {Hanke}},\ }\bibfield  {title} {\enquote
  {\bibinfo {title} {Model reduction techniques for the computation of extended
  {{Markov}} parameterizations for generalized {{Langevin}} equations},}\
  }\href {https://doi.org/10.1088/1361-648X/abe6df} {\bibfield  {journal}
  {\bibinfo  {journal} {Journal of Physics: Condensed Matter}\ }\textbf
  {\bibinfo {volume} {33}},\ \bibinfo {pages} {214003} (\bibinfo {year}
  {2021})}\BibitemShut {NoStop}%
\bibitem [{\citenamefont {Daldrop}, \citenamefont {Kowalik},\ and\
  \citenamefont {Netz}(2017)}]{daldropExternalPotentialModifies2017}%
  \BibitemOpen
  \bibfield  {author} {\bibinfo {author} {\bibfnamefont {J.~O.}\ \bibnamefont
  {Daldrop}}, \bibinfo {author} {\bibfnamefont {B.~G.}\ \bibnamefont
  {Kowalik}},\ and\ \bibinfo {author} {\bibfnamefont {R.~R.}\ \bibnamefont
  {Netz}},\ }\bibfield  {title} {\enquote {\bibinfo {title} {External potential
  modifies friction of molecular solutes in water},}\ }\href
  {https://doi.org/10.1103/PhysRevX.7.041065} {\bibfield  {journal} {\bibinfo
  {journal} {Physical Review X}\ }\textbf {\bibinfo {volume} {7}},\ \bibinfo
  {pages} {041065} (\bibinfo {year} {2017})}\BibitemShut {NoStop}%
\bibitem [{\citenamefont {Ma}, \citenamefont {Li},\ and\ \citenamefont
  {Liu}(2016)}]{ma2016derivation}%
  \BibitemOpen
  \bibfield  {author} {\bibinfo {author} {\bibfnamefont {L.}~\bibnamefont
  {Ma}}, \bibinfo {author} {\bibfnamefont {X.}~\bibnamefont {Li}},\ and\
  \bibinfo {author} {\bibfnamefont {C.}~\bibnamefont {Liu}},\ }\bibfield
  {title} {\enquote {\bibinfo {title} {The derivation and approximation of
  coarse-grained dynamics from langevin dynamics},}\ }\href@noop {} {\bibfield
  {journal} {\bibinfo  {journal} {The Journal of chemical physics}\ }\textbf
  {\bibinfo {volume} {145}} (\bibinfo {year} {2016})}\BibitemShut {NoStop}%
\bibitem [{\citenamefont {Grogan}\ \emph {et~al.}(2020)\citenamefont {Grogan},
  \citenamefont {Lei}, \citenamefont {Li},\ and\ \citenamefont
  {Baker}}]{groganDatadrivenMolecularModeling2020}%
  \BibitemOpen
  \bibfield  {author} {\bibinfo {author} {\bibfnamefont {F.}~\bibnamefont
  {Grogan}}, \bibinfo {author} {\bibfnamefont {H.}~\bibnamefont {Lei}},
  \bibinfo {author} {\bibfnamefont {X.}~\bibnamefont {Li}},\ and\ \bibinfo
  {author} {\bibfnamefont {N.~A.}\ \bibnamefont {Baker}},\ }\bibfield  {title}
  {\enquote {\bibinfo {title} {Data-driven molecular modeling with the
  generalized {{Langevin}} equation},}\ }\href
  {https://doi.org/10.1016/j.jcp.2020.109633} {\bibfield  {journal} {\bibinfo
  {journal} {Journal of Computational Physics}\ }\textbf {\bibinfo {volume}
  {418}},\ \bibinfo {pages} {109633} (\bibinfo {year} {2020})}\BibitemShut
  {NoStop}%
\bibitem [{\citenamefont {Ayaz}\ \emph {et~al.}(2021)\citenamefont {Ayaz},
  \citenamefont {Tepper}, \citenamefont {Br{\"u}nig}, \citenamefont {Kappler},
  \citenamefont {Daldrop},\ and\ \citenamefont
  {Netz}}]{ayazNonMarkovianModelingProtein2021}%
  \BibitemOpen
  \bibfield  {author} {\bibinfo {author} {\bibfnamefont {C.}~\bibnamefont
  {Ayaz}}, \bibinfo {author} {\bibfnamefont {L.}~\bibnamefont {Tepper}},
  \bibinfo {author} {\bibfnamefont {F.~N.}\ \bibnamefont {Br{\"u}nig}},
  \bibinfo {author} {\bibfnamefont {J.}~\bibnamefont {Kappler}}, \bibinfo
  {author} {\bibfnamefont {J.~O.}\ \bibnamefont {Daldrop}},\ and\ \bibinfo
  {author} {\bibfnamefont {R.~R.}\ \bibnamefont {Netz}},\ }\bibfield  {title}
  {\enquote {\bibinfo {title} {Non-markovian modeling of protein folding},}\
  }\href@noop {} {\bibfield  {journal} {\bibinfo  {journal} {Proceedings of the
  National Academy of Sciences}\ }\textbf {\bibinfo {volume} {118}},\ \bibinfo
  {pages} {e2023856118} (\bibinfo {year} {2021})}\BibitemShut {NoStop}%
\bibitem [{\citenamefont {Dalton}\ \emph {et~al.}(2023)\citenamefont {Dalton},
  \citenamefont {Ayaz}, \citenamefont {Kiefer}, \citenamefont {Klimek},
  \citenamefont {Tepper},\ and\ \citenamefont
  {Netz}}]{daltonFastProteinFolding2023}%
  \BibitemOpen
  \bibfield  {author} {\bibinfo {author} {\bibfnamefont {B.~A.}\ \bibnamefont
  {Dalton}}, \bibinfo {author} {\bibfnamefont {C.}~\bibnamefont {Ayaz}},
  \bibinfo {author} {\bibfnamefont {H.}~\bibnamefont {Kiefer}}, \bibinfo
  {author} {\bibfnamefont {A.}~\bibnamefont {Klimek}}, \bibinfo {author}
  {\bibfnamefont {L.}~\bibnamefont {Tepper}},\ and\ \bibinfo {author}
  {\bibfnamefont {R.~R.}\ \bibnamefont {Netz}},\ }\bibfield  {title} {\enquote
  {\bibinfo {title} {Fast protein folding is governed by memory-dependent
  friction},}\ }\href@noop {} {\bibfield  {journal} {\bibinfo  {journal}
  {Proceedings of the National Academy of Sciences}\ }\textbf {\bibinfo
  {volume} {120}},\ \bibinfo {pages} {e2220068120} (\bibinfo {year}
  {2023})}\BibitemShut {NoStop}%
\bibitem [{\citenamefont {She}, \citenamefont {Ge},\ and\ \citenamefont
  {Lei}(2023)}]{sheDatadrivenConstructionStochastic2023}%
  \BibitemOpen
  \bibfield  {author} {\bibinfo {author} {\bibfnamefont {Z.}~\bibnamefont
  {She}}, \bibinfo {author} {\bibfnamefont {P.}~\bibnamefont {Ge}},\ and\
  \bibinfo {author} {\bibfnamefont {H.}~\bibnamefont {Lei}},\ }\bibfield
  {title} {\enquote {\bibinfo {title} {Data-driven construction of stochastic
  reduced dynamics encoded with non-{{Markovian}} features},}\ }\href
  {https://doi.org/10.1063/5.0130033} {\bibfield  {journal} {\bibinfo
  {journal} {The Journal of Chemical Physics}\ }\textbf {\bibinfo {volume}
  {158}},\ \bibinfo {pages} {034102} (\bibinfo {year} {2023})}\BibitemShut
  {NoStop}%
\bibitem [{\citenamefont {Yoshimoto}\ \emph {et~al.}(2017)\citenamefont
  {Yoshimoto}, \citenamefont {Li}, \citenamefont {Kinefuchi},\ and\
  \citenamefont
  {Karniadakis}}]{yoshimotoConstructionNonMarkovianCoarsegrained2017}%
  \BibitemOpen
  \bibfield  {author} {\bibinfo {author} {\bibfnamefont {Y.}~\bibnamefont
  {Yoshimoto}}, \bibinfo {author} {\bibfnamefont {Z.}~\bibnamefont {Li}},
  \bibinfo {author} {\bibfnamefont {I.}~\bibnamefont {Kinefuchi}},\ and\
  \bibinfo {author} {\bibfnamefont {G.~E.}\ \bibnamefont {Karniadakis}},\
  }\bibfield  {title} {\enquote {\bibinfo {title} {Construction of
  non-{{Markovian}} coarse-grained models employing the
  {{Mori}}{\textendash}{{Zwanzig}} formalism and iterative {{Boltzmann}}
  inversion},}\ }\href {https://doi.org/10.1063/1.5009041} {\bibfield
  {journal} {\bibinfo  {journal} {The Journal of Chemical Physics}\ }\textbf
  {\bibinfo {volume} {147}},\ \bibinfo {pages} {244110} (\bibinfo {year}
  {2017})}\BibitemShut {NoStop}%
\bibitem [{\citenamefont {Virtanen}\ \emph {et~al.}(2020)\citenamefont
  {Virtanen}, \citenamefont {Gommers}, \citenamefont {Oliphant}, \citenamefont
  {Haberland}, \citenamefont {Reddy}, \citenamefont {Cournapeau}, \citenamefont
  {Burovski}, \citenamefont {Peterson}, \citenamefont {Weckesser},
  \citenamefont {Bright}, \citenamefont {{van der Walt}}, \citenamefont
  {Brett}, \citenamefont {Wilson}, \citenamefont {Millman}, \citenamefont
  {Mayorov}, \citenamefont {Nelson}, \citenamefont {Jones}, \citenamefont
  {Kern}, \citenamefont {Larson}, \citenamefont {Carey}, \citenamefont {Polat},
  \citenamefont {Feng}, \citenamefont {Moore}, \citenamefont {VanderPlas},
  \citenamefont {Laxalde}, \citenamefont {Perktold}, \citenamefont {Cimrman},
  \citenamefont {Henriksen}, \citenamefont {Quintero}, \citenamefont {Harris},
  \citenamefont {Archibald}, \citenamefont {Ribeiro}, \citenamefont
  {Pedregosa},\ and\ \citenamefont {{van
  Mulbregt}}}]{virtanenSciPyFundamentalAlgorithms2020}%
  \BibitemOpen
  \bibfield  {author} {\bibinfo {author} {\bibfnamefont {P.}~\bibnamefont
  {Virtanen}}, \bibinfo {author} {\bibfnamefont {R.}~\bibnamefont {Gommers}},
  \bibinfo {author} {\bibfnamefont {T.~E.}\ \bibnamefont {Oliphant}}, \bibinfo
  {author} {\bibfnamefont {M.}~\bibnamefont {Haberland}}, \bibinfo {author}
  {\bibfnamefont {T.}~\bibnamefont {Reddy}}, \bibinfo {author} {\bibfnamefont
  {D.}~\bibnamefont {Cournapeau}}, \bibinfo {author} {\bibfnamefont
  {E.}~\bibnamefont {Burovski}}, \bibinfo {author} {\bibfnamefont
  {P.}~\bibnamefont {Peterson}}, \bibinfo {author} {\bibfnamefont
  {W.}~\bibnamefont {Weckesser}}, \bibinfo {author} {\bibfnamefont
  {J.}~\bibnamefont {Bright}}, \bibinfo {author} {\bibfnamefont {S.~J.}\
  \bibnamefont {{van der Walt}}}, \bibinfo {author} {\bibfnamefont
  {M.}~\bibnamefont {Brett}}, \bibinfo {author} {\bibfnamefont
  {J.}~\bibnamefont {Wilson}}, \bibinfo {author} {\bibfnamefont {K.~J.}\
  \bibnamefont {Millman}}, \bibinfo {author} {\bibfnamefont {N.}~\bibnamefont
  {Mayorov}}, \bibinfo {author} {\bibfnamefont {A.~R.~J.}\ \bibnamefont
  {Nelson}}, \bibinfo {author} {\bibfnamefont {E.}~\bibnamefont {Jones}},
  \bibinfo {author} {\bibfnamefont {R.}~\bibnamefont {Kern}}, \bibinfo {author}
  {\bibfnamefont {E.}~\bibnamefont {Larson}}, \bibinfo {author} {\bibfnamefont
  {C.~J.}\ \bibnamefont {Carey}}, \bibinfo {author} {\bibfnamefont
  {{\.I}.}~\bibnamefont {Polat}}, \bibinfo {author} {\bibfnamefont
  {Y.}~\bibnamefont {Feng}}, \bibinfo {author} {\bibfnamefont {E.~W.}\
  \bibnamefont {Moore}}, \bibinfo {author} {\bibfnamefont {J.}~\bibnamefont
  {VanderPlas}}, \bibinfo {author} {\bibfnamefont {D.}~\bibnamefont {Laxalde}},
  \bibinfo {author} {\bibfnamefont {J.}~\bibnamefont {Perktold}}, \bibinfo
  {author} {\bibfnamefont {R.}~\bibnamefont {Cimrman}}, \bibinfo {author}
  {\bibfnamefont {I.}~\bibnamefont {Henriksen}}, \bibinfo {author}
  {\bibfnamefont {E.~A.}\ \bibnamefont {Quintero}}, \bibinfo {author}
  {\bibfnamefont {C.~R.}\ \bibnamefont {Harris}}, \bibinfo {author}
  {\bibfnamefont {A.~M.}\ \bibnamefont {Archibald}}, \bibinfo {author}
  {\bibfnamefont {A.~H.}\ \bibnamefont {Ribeiro}}, \bibinfo {author}
  {\bibfnamefont {F.}~\bibnamefont {Pedregosa}},\ and\ \bibinfo {author}
  {\bibfnamefont {P.}~\bibnamefont {{van Mulbregt}}},\ }\bibfield  {title}
  {\enquote {\bibinfo {title} {{{SciPy}} 1.0: Fundamental algorithms for
  scientific computing in {{Python}}},}\ }\href
  {https://doi.org/10.1038/s41592-019-0686-2} {\bibfield  {journal} {\bibinfo
  {journal} {Nature Methods}\ }\textbf {\bibinfo {volume} {17}},\ \bibinfo
  {pages} {261--272} (\bibinfo {year} {2020})}\BibitemShut {NoStop}%
\bibitem [{\citenamefont {Ayaz}\ \emph {et~al.}(2022)\citenamefont {Ayaz},
  \citenamefont {Scalfi}, \citenamefont {Dalton},\ and\ \citenamefont
  {Netz}}]{ayazGeneralizedLangevinEquation2022}%
  \BibitemOpen
  \bibfield  {author} {\bibinfo {author} {\bibfnamefont {C.}~\bibnamefont
  {Ayaz}}, \bibinfo {author} {\bibfnamefont {L.}~\bibnamefont {Scalfi}},
  \bibinfo {author} {\bibfnamefont {B.~A.}\ \bibnamefont {Dalton}},\ and\
  \bibinfo {author} {\bibfnamefont {R.~R.}\ \bibnamefont {Netz}},\ }\bibfield
  {title} {\enquote {\bibinfo {title} {Generalized {{Langevin}} equation with a
  nonlinear potential of mean force and nonlinear memory friction from a hybrid
  projection scheme},}\ }\href {https://doi.org/10.1103/PhysRevE.105.054138}
  {\bibfield  {journal} {\bibinfo  {journal} {Physical Review E}\ }\textbf
  {\bibinfo {volume} {105}},\ \bibinfo {pages} {054138} (\bibinfo {year}
  {2022})}\BibitemShut {NoStop}%
\bibitem [{\citenamefont {Vroylandt}\ and\ \citenamefont
  {Monmarch{\'e}}(2022)}]{vroylandtPositiondependentMemoryKernel2022}%
  \BibitemOpen
  \bibfield  {author} {\bibinfo {author} {\bibfnamefont {H.}~\bibnamefont
  {Vroylandt}}\ and\ \bibinfo {author} {\bibfnamefont {P.}~\bibnamefont
  {Monmarch{\'e}}},\ }\bibfield  {title} {\enquote {\bibinfo {title}
  {Position-dependent memory kernel in generalized {{Langevin}} equations:
  {{Theory}} and numerical estimation},}\ }\href
  {https://doi.org/10.1063/5.0094566} {\bibfield  {journal} {\bibinfo
  {journal} {The Journal of Chemical Physics}\ }\textbf {\bibinfo {volume}
  {156}},\ \bibinfo {pages} {244105} (\bibinfo {year} {2022})}\BibitemShut
  {NoStop}%
\bibitem [{\citenamefont {Lyu}\ and\ \citenamefont
  {Lei}(2023)}]{lyuConstructionCoarseGrainedMolecular2023}%
  \BibitemOpen
  \bibfield  {author} {\bibinfo {author} {\bibfnamefont {L.}~\bibnamefont
  {Lyu}}\ and\ \bibinfo {author} {\bibfnamefont {H.}~\bibnamefont {Lei}},\
  }\bibfield  {title} {\enquote {\bibinfo {title} {Construction of
  {{Coarse-Grained Molecular Dynamics}} with {{Many-Body Non-Markovian
  Memory}}},}\ }\href {https://doi.org/10.1103/PhysRevLett.131.177301}
  {\bibfield  {journal} {\bibinfo  {journal} {Physical Review Letters}\
  }\textbf {\bibinfo {volume} {131}},\ \bibinfo {pages} {177301} (\bibinfo
  {year} {2023})}\BibitemShut {NoStop}%
\bibitem [{\citenamefont {Pastorino}\ \emph {et~al.}(2007)\citenamefont
  {Pastorino}, \citenamefont {Kreer}, \citenamefont {M{\"u}ller},\ and\
  \citenamefont {Binder}}]{pastorinoComparisonDissipativeParticle2007}%
  \BibitemOpen
  \bibfield  {author} {\bibinfo {author} {\bibfnamefont {C.}~\bibnamefont
  {Pastorino}}, \bibinfo {author} {\bibfnamefont {T.}~\bibnamefont {Kreer}},
  \bibinfo {author} {\bibfnamefont {M.}~\bibnamefont {M{\"u}ller}},\ and\
  \bibinfo {author} {\bibfnamefont {K.}~\bibnamefont {Binder}},\ }\bibfield
  {title} {\enquote {\bibinfo {title} {Comparison of dissipative particle
  dynamics and {{Langevin}} thermostats for out-of-equilibrium simulations of
  polymeric systems},}\ }\href {https://doi.org/10.1103/PhysRevE.76.026706}
  {\bibfield  {journal} {\bibinfo  {journal} {Physical Review E}\ }\textbf
  {\bibinfo {volume} {76}},\ \bibinfo {pages} {026706} (\bibinfo {year}
  {2007})}\BibitemShut {NoStop}%
\bibitem [{\citenamefont {Yoshimoto}\ \emph {et~al.}(2013)\citenamefont
  {Yoshimoto}, \citenamefont {Kinefuchi}, \citenamefont {Mima}, \citenamefont
  {Fukushima}, \citenamefont {Tokumasu},\ and\ \citenamefont
  {Takagi}}]{yoshimotoBottomupConstructionInteraction2013}%
  \BibitemOpen
  \bibfield  {author} {\bibinfo {author} {\bibfnamefont {Y.}~\bibnamefont
  {Yoshimoto}}, \bibinfo {author} {\bibfnamefont {I.}~\bibnamefont
  {Kinefuchi}}, \bibinfo {author} {\bibfnamefont {T.}~\bibnamefont {Mima}},
  \bibinfo {author} {\bibfnamefont {A.}~\bibnamefont {Fukushima}}, \bibinfo
  {author} {\bibfnamefont {T.}~\bibnamefont {Tokumasu}},\ and\ \bibinfo
  {author} {\bibfnamefont {S.}~\bibnamefont {Takagi}},\ }\bibfield  {title}
  {\enquote {\bibinfo {title} {Bottom-up construction of interaction models of
  non-{{Markovian}} dissipative particle dynamics},}\ }\href
  {https://doi.org/10.1103/PhysRevE.88.043305} {\bibfield  {journal} {\bibinfo
  {journal} {Physical Review E}\ }\textbf {\bibinfo {volume} {88}},\ \bibinfo
  {pages} {043305} (\bibinfo {year} {2013})}\BibitemShut {NoStop}%
\bibitem [{\citenamefont {Thompson}\ \emph {et~al.}(2022)\citenamefont
  {Thompson}, \citenamefont {Aktulga}, \citenamefont {Berger}, \citenamefont
  {Bolintineanu}, \citenamefont {Brown}, \citenamefont {Crozier}, \citenamefont
  {{in 't Veld}}, \citenamefont {Kohlmeyer}, \citenamefont {Moore},
  \citenamefont {Nguyen}, \citenamefont {Shan}, \citenamefont {Stevens},
  \citenamefont {Tranchida}, \citenamefont {Trott},\ and\ \citenamefont
  {Plimpton}}]{thompsonLAMMPSFlexibleSimulation2022}%
  \BibitemOpen
  \bibfield  {author} {\bibinfo {author} {\bibfnamefont {A.~P.}\ \bibnamefont
  {Thompson}}, \bibinfo {author} {\bibfnamefont {H.~M.}\ \bibnamefont
  {Aktulga}}, \bibinfo {author} {\bibfnamefont {R.}~\bibnamefont {Berger}},
  \bibinfo {author} {\bibfnamefont {D.~S.}\ \bibnamefont {Bolintineanu}},
  \bibinfo {author} {\bibfnamefont {W.~M.}\ \bibnamefont {Brown}}, \bibinfo
  {author} {\bibfnamefont {P.~S.}\ \bibnamefont {Crozier}}, \bibinfo {author}
  {\bibfnamefont {P.~J.}\ \bibnamefont {{in 't Veld}}}, \bibinfo {author}
  {\bibfnamefont {A.}~\bibnamefont {Kohlmeyer}}, \bibinfo {author}
  {\bibfnamefont {S.~G.}\ \bibnamefont {Moore}}, \bibinfo {author}
  {\bibfnamefont {T.~D.}\ \bibnamefont {Nguyen}}, \bibinfo {author}
  {\bibfnamefont {R.}~\bibnamefont {Shan}}, \bibinfo {author} {\bibfnamefont
  {M.~J.}\ \bibnamefont {Stevens}}, \bibinfo {author} {\bibfnamefont
  {J.}~\bibnamefont {Tranchida}}, \bibinfo {author} {\bibfnamefont
  {C.}~\bibnamefont {Trott}},\ and\ \bibinfo {author} {\bibfnamefont {S.~J.}\
  \bibnamefont {Plimpton}},\ }\bibfield  {title} {\enquote {\bibinfo {title}
  {{{LAMMPS}} - a flexible simulation tool for particle-based materials
  modeling at the atomic, meso, and continuum scales},}\ }\href
  {https://doi.org/10.1016/j.cpc.2021.108171} {\bibfield  {journal} {\bibinfo
  {journal} {Computer Physics Communications}\ }\textbf {\bibinfo {volume}
  {271}},\ \bibinfo {pages} {108171} (\bibinfo {year} {2022})}\BibitemShut
  {NoStop}%
\bibitem [{\citenamefont {Jorgensen}, \citenamefont {Maxwell},\ and\
  \citenamefont {{Tirado-Rives}}(1996)}]{jorgensenDevelopmentTestingOPLS1996}%
  \BibitemOpen
  \bibfield  {author} {\bibinfo {author} {\bibfnamefont {W.~L.}\ \bibnamefont
  {Jorgensen}}, \bibinfo {author} {\bibfnamefont {D.~S.}\ \bibnamefont
  {Maxwell}},\ and\ \bibinfo {author} {\bibfnamefont {J.}~\bibnamefont
  {{Tirado-Rives}}},\ }\bibfield  {title} {\enquote {\bibinfo {title}
  {Development and {{Testing}} of the {{OPLS All-Atom Force Field}} on
  {{Conformational Energetics}} and {{Properties}} of {{Organic Liquids}}},}\
  }\href {https://doi.org/10.1021/ja9621760} {\bibfield  {journal} {\bibinfo
  {journal} {Journal of the American Chemical Society}\ }\textbf {\bibinfo
  {volume} {118}},\ \bibinfo {pages} {11225--11236} (\bibinfo {year}
  {1996})}\BibitemShut {NoStop}%
\bibitem [{\citenamefont {Rackers}\ \emph {et~al.}(2018)\citenamefont
  {Rackers}, \citenamefont {Wang}, \citenamefont {Lu}, \citenamefont {Laury},
  \citenamefont {Lagard{\`e}re}, \citenamefont {Schnieders}, \citenamefont
  {Piquemal}, \citenamefont {Ren},\ and\ \citenamefont
  {Ponder}}]{rackersTinkerSoftwareTools2018}%
  \BibitemOpen
  \bibfield  {author} {\bibinfo {author} {\bibfnamefont {J.~A.}\ \bibnamefont
  {Rackers}}, \bibinfo {author} {\bibfnamefont {Z.}~\bibnamefont {Wang}},
  \bibinfo {author} {\bibfnamefont {C.}~\bibnamefont {Lu}}, \bibinfo {author}
  {\bibfnamefont {M.~L.}\ \bibnamefont {Laury}}, \bibinfo {author}
  {\bibfnamefont {L.}~\bibnamefont {Lagard{\`e}re}}, \bibinfo {author}
  {\bibfnamefont {M.~J.}\ \bibnamefont {Schnieders}}, \bibinfo {author}
  {\bibfnamefont {J.-P.}\ \bibnamefont {Piquemal}}, \bibinfo {author}
  {\bibfnamefont {P.}~\bibnamefont {Ren}},\ and\ \bibinfo {author}
  {\bibfnamefont {J.~W.}\ \bibnamefont {Ponder}},\ }\bibfield  {title}
  {\enquote {\bibinfo {title} {Tinker 8: {{Software Tools}} for {{Molecular
  Design}}},}\ }\href {https://doi.org/10.1021/acs.jctc.8b00529} {\bibfield
  {journal} {\bibinfo  {journal} {Journal of Chemical Theory and Computation}\
  }\textbf {\bibinfo {volume} {14}},\ \bibinfo {pages} {5273--5289} (\bibinfo
  {year} {2018})}\BibitemShut {NoStop}%
\bibitem [{\citenamefont {Jewett}\ \emph {et~al.}(2021)\citenamefont {Jewett},
  \citenamefont {Stelter}, \citenamefont {Lambert}, \citenamefont {Saladi},
  \citenamefont {Roscioni}, \citenamefont {Ricci}, \citenamefont {Autin},
  \citenamefont {Maritan}, \citenamefont {Bashusqeh}, \citenamefont {Keyes}
  \emph {et~al.}}]{jewettMoltemplateToolCoarseGrained2021}%
  \BibitemOpen
  \bibfield  {author} {\bibinfo {author} {\bibfnamefont {A.~I.}\ \bibnamefont
  {Jewett}}, \bibinfo {author} {\bibfnamefont {D.}~\bibnamefont {Stelter}},
  \bibinfo {author} {\bibfnamefont {J.}~\bibnamefont {Lambert}}, \bibinfo
  {author} {\bibfnamefont {S.~M.}\ \bibnamefont {Saladi}}, \bibinfo {author}
  {\bibfnamefont {O.~M.}\ \bibnamefont {Roscioni}}, \bibinfo {author}
  {\bibfnamefont {M.}~\bibnamefont {Ricci}}, \bibinfo {author} {\bibfnamefont
  {L.}~\bibnamefont {Autin}}, \bibinfo {author} {\bibfnamefont
  {M.}~\bibnamefont {Maritan}}, \bibinfo {author} {\bibfnamefont {S.~M.}\
  \bibnamefont {Bashusqeh}}, \bibinfo {author} {\bibfnamefont {T.}~\bibnamefont
  {Keyes}}, \emph {et~al.},\ }\bibfield  {title} {\enquote {\bibinfo {title}
  {Moltemplate: A tool for coarse-grained modeling of complex biological matter
  and soft condensed matter physics},}\ }\href@noop {} {\bibfield  {journal}
  {\bibinfo  {journal} {Journal of molecular biology}\ }\textbf {\bibinfo
  {volume} {433}},\ \bibinfo {pages} {166841} (\bibinfo {year}
  {2021})}\BibitemShut {NoStop}%
\bibitem [{\citenamefont {R{\"u}hle}\ \emph {et~al.}(2009)\citenamefont
  {R{\"u}hle}, \citenamefont {Junghans}, \citenamefont {Lukyanov},
  \citenamefont {Kremer},\ and\ \citenamefont
  {Andrienko}}]{ruhleVersatileObjectOrientedToolkit2009}%
  \BibitemOpen
  \bibfield  {author} {\bibinfo {author} {\bibfnamefont {V.}~\bibnamefont
  {R{\"u}hle}}, \bibinfo {author} {\bibfnamefont {C.}~\bibnamefont {Junghans}},
  \bibinfo {author} {\bibfnamefont {A.}~\bibnamefont {Lukyanov}}, \bibinfo
  {author} {\bibfnamefont {K.}~\bibnamefont {Kremer}},\ and\ \bibinfo {author}
  {\bibfnamefont {D.}~\bibnamefont {Andrienko}},\ }\bibfield  {title} {\enquote
  {\bibinfo {title} {Versatile {{Object-Oriented Toolkit}} for
  {{Coarse-Graining Applications}}},}\ }\href
  {https://doi.org/10.1021/ct900369w} {\bibfield  {journal} {\bibinfo
  {journal} {Journal of Chemical Theory and Computation}\ }\textbf {\bibinfo
  {volume} {5}},\ \bibinfo {pages} {3211--3223} (\bibinfo {year}
  {2009})}\BibitemShut {NoStop}%
\bibitem [{\citenamefont
  {Ceriotti}(2010)}]{ceriottiNovelFrameworkEnhanced2010}%
  \BibitemOpen
  \bibfield  {author} {\bibinfo {author} {\bibfnamefont {M.}~\bibnamefont
  {Ceriotti}},\ }\emph {\bibinfo {title} {A Novel Framework for Enhanced
  Molecular Dynamics Based on the Generalized {{Langevin}} Equation}},\
  \href@noop {} {Ph.D. thesis},\ \bibinfo  {school} {ETH Zurich} (\bibinfo
  {year} {2010})\BibitemShut {NoStop}%
\end{thebibliography}%
\clearpage
\appendix
\label{app}
\section{Computational details}
\label{app:comp}
\subsection{Atomistic ethanol simulations}
\label{app:ethanol}
The FG atomistic reference simulations of liquid ethanol were performed with LAMMPS\cite{thompsonLAMMPSFlexibleSimulation2022} using the OPLS/AA force field.\cite{jorgensenDevelopmentTestingOPLS1996,rackersTinkerSoftwareTools2018} A cutoff of 1.1 nm was applied for Lennard–Jones and real-space electrostatic interactions while long-range electrostatics were treated by the particle-particle particle-mesh method.   
We used MolTemplate\cite{jewettMoltemplateToolCoarseGrained2021} to generate an initial configuration of 2744 ethanol molecules. 
All FG simulations were carried out with a time step of 0.5 fs.
An energy minimized structure was equilibrated under NVT conditions at 300 K with a Nos\'e-Hoover thermostat using a time constant of 50 fs. The simulation box size was then relaxed under NPT conditions at a pressure of 1 atm using the Nos\'e-Hoover barostat with a time constant of 500 fs. A tail-correction was applied for pressure calculations. The average box length was evaluated using an additional NPT run of 200 ps and set to  6.42778 nm for all other simulations. 
A final NVT equilibration run was performed using a Langevin thermostat with a time constant of 500 fs to generate an equilibrated structure for the production run.

To calculate the center of mass (COM) radial distribution function (RDF) 
a 0.5 ns trajectory was generated using a Langevin thermostat with a time constant of 500 fs. The COM RDF was computed using the VOTCA package.~\cite{ruhleVersatileObjectOrientedToolkit2009} 

In addition, ten snapshots were taken to be used as initial states for independent runs to generate reference dynamic data, i.e., the VACF.
From these ten initial snapshots, independent NVT trajectories were generated using the Nose-Hoover thermostat with a time constant of 2.5 ps and a length of 300 ps each. Frames were saved every 10 fs. From these runs, ten independent VACFs were calculated and averaged to serve as a reference (Fig.~\ref{fig:GN_results} b) and e)). The target integrated memory kernel $G^\text{tgt}(t)$ was derived accordingly from Eq.~\ref{eq:volterra}.
\subsection{Coarse-grained conservative interactions}
\label{app:ethanol_ibi}
\begin{figure}
    \centering
    \includegraphics{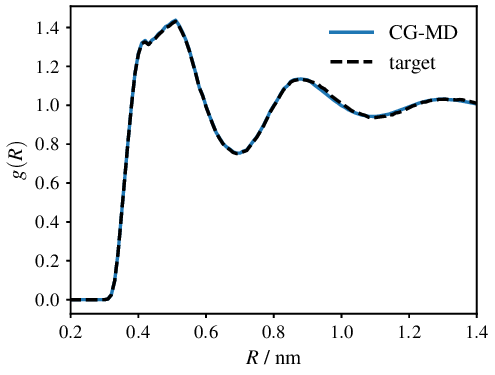}
    \caption{COM radial distribution function ($g(R)$) of ethanol from FG-MD and CG-MD using the IBI potential. }
    \label{appfig:rdf}
\end{figure}

For the CG representation, all ethanol molecules are mapped onto their respective COM, reducing the DoF by a factor of 9. The RDF from the mapped FG trajectory was used as the target to generate a CG potential using the IBI\cite{reithDerivingEffectiveMesoscale2003} method as implemented in VOTCA.\cite{ruhleVersatileObjectOrientedToolkit2009}.
Each IBI iteration was started from the same initial structure of a mapped equilibrated snapshot of the FG NVT system. We used tabulated potentials with a cutoff of 0.9 nm, a time step of 2 fs, and a Langevin thermostat with a time constant of 0.2 ps. In each iteration, velocities were initialized according to the Maxwell-Boltzmann distribution at 300 K. The system was equilibrated for 40 ps, after which every 100th frame of a 200 ps simulation was saved. 
The IBI method was run for 300 iterations. The COM RDF ($g(R)$) of the FG and the final IBI model are compared in Fig.~\ref{appfig:rdf}.
\subsection{Coarse-grained simulations without a dissipative thermostat}
\label{app:ethanol_cg}
We performed standard CG-MD simulations with the previously derived IBI potential to evaluate $G^\circ(t)$, the integrated single-particle memory kernel of the Hamiltonian CG model. Velocities were initialized according to the Maxwell-Boltzmann distribution at 300 K. The system was equilibrated for 100 ps using a Langevin thermostat with a time constant of 0.2 ps. 
A Nose-Hoover thermostat with a time constant of 4 ps was used for production. The trajectories were generated by $10\times 400$ ps simulations, saving frames only every 10 fs after the first 100 ps. By skipping the first 100 ps in each subtrajectory, we generate sufficiently independent trajectories to ensure proper convergence of the subsequently computed VACF ($C_{VV}^\circ(t)$).

\subsection{Coarse-grained simulations within IOMK and IOMK-GN iterations}
Coarse-grained aux-GLE simulations for IOMK and IOMK-GN iterations were performed using the aux-GLE\cite{ceriottiColoredNoiseThermostatsCarte2010a} implementation in LAMMPS.\cite{thompsonLAMMPSFlexibleSimulation2022}
Except for the thermostat, the setup was chosen equivalent to Sec.~\ref{app:ethanol_cg}.
VACFs were generated from  slightly shorter and fewer ($5\times 300$ ps) subtrajectories. 

\section{Numerical considerations}
\label{app:num}
\subsection{Update optimization by varying regularization} 
\label{app:step}
A common approach to improve the performance of iterative methods is to optimize the iteration step. For example, a line search along the $\boldsymbol{\Delta}^{i+1}$ direction specified by the initial scheme can be used to find the parameters that locally minimize the residual along the update direction.\cite{reinhardtNichtlineareOptimierung2013} We, however, choose an alternative approach, where the regularization parameter $\alpha$ and, therefore, both the norm and the direction of the iteration step are optimized simultaneously. 

To this end, consider a function $\hat{\bm{f}}(\bm{x})\approx \bm{f}(\bm{x})$. 
Fixing the maximum value of $\alpha$ to $\alpha_{\text{max}}$, we can compute a Gauss-Newton step $\boldsymbol{\Delta}^{i+1}(\epsilon)$ for a set of scaling parameters $0 < \epsilon\leq 1$ with $\alpha = \epsilon \alpha_{\text{max}}$ and choose the actual step such that 
\begin{equation}
    \min_{0<\epsilon\leq 1} \norm{\hat{\bm{f}}(\bm{x}^i+\boldsymbol{\Delta}^{i+1}(\epsilon))-\bm{f}^{\text{tgt}}}^2.
    \label{eq:min_s}
\end{equation}
This is a computationally cheap way to prevent an \textit{a priori} chosen regularization parameter from strongly slowing down convergence. We chose  $\epsilon<1$, such that the size of the update step can only be increased. While this may cause some iterations to increase the residuals, it prevents the procedure from getting stuck in regions where progress would be slow with stronger regularization.
As long as $\alpha_{\text{max}}$ is chosen such that $\epsilon=1$ would yield stable iterations, this strategy is expected to speed up convergence. However, we note that consequently, $||\bm{r}||^2$ does not necessarily decrease with every iteration, which can be an advantage as local minima of $||\bm{r}||^2$ can be overcome.
\subsection{Normalization of memory kernels} 
\label{app:norm}
To optimize $\bm{A}$, a reasonable initial guess and appropriate bounds on the parameter space should be used.
The choices fore the initial guess and the bounds are not transferable. To ensure that a single heuristic choice (described in Appendix \ref{app:bounds} and \ref{app:guess}) applies to a wide range of systems, we propose  to normalize the target memory kernel.
From that we obtain a normalized drift matrix that must be renormalized to parameterize the thermostat.

To understand the normalization process, consider a discretized memory kernel  $G\in \mathbb{R}^\mathcal{M}$, equally spaced in time such that $t^{\text{max}}=\mathcal{M}\Delta t$ is the time up to which the memory kernel is considered in the optimization process. 
We define a normalized time-axis as  $t_m^{\text{norm}}= \frac{t_{m}}{t^{\text{max}}}$.
Similarly, the memory kernel is normalized to its maximum value as  $G^{\text{max}} = \text{max}\left(\bm{G}\right)$, and the normalized memory kernel is defined as $\bm{G}^{\text{norm}} = \frac{\bm{G}}{G^{\text{max}}}$. 
The optimization is performed on normalized data, and we extract the normalized drift matrix $\bm{A}^{\text{norm}}$ from which $\bm{A}$ can be determined by
\begin{equation}
    \bm{A} =  \begin{pmatrix}
        0 &  \sqrt{\frac{G^{\text{max}}}{t^{\text{max}}}}\left(\bm{A}^{\text{norm}}_{Ps}\right)^T\\
        -\sqrt{\frac{G^{\text{max}}}{t^{\text{max}}}}\bm{A}^{\text{norm}}_{Ps}& \bm{A}^{\text{norm}}_{ss}/t^{\text{max}}
    \end{pmatrix}.
    \label{eq:A_mat_renorm}
\end{equation}
\subsection{Bounds}
\label{app:bounds}
The parameters making up the drift matrix $\bm{A}$ are restrained by the FDT which, with Eq.~\ref{eq:A_mat2}, requires that $\left(\bm{A}_{ss}\right)_{kk} \geq 0$ for all $k$. 
Other than that, the parameters in the drift matrix only have to be in $\mathbb{R} \backslash \{0\}$ to guarantee $\lim_{t\rightarrow\infty}G(t) > 0$.

To increase numerical stability, we chose to restrict the parameter space further.
Changing the sign does not affect the memory kernel for all off-diagonal parameters in $\bm{A}$.
Still, we want to prevent sign reversals since an update that attempts to reduce an entry $(\bm{A}^i)_{kl}$ might otherwise result in $|(\bm{A}^{i+1})_{kl}|>(\bm{A}^i)_{kl}$.
Consequently, we chose all the parameters to be in $\mathbb{R}_+$. 
Furthermore, when using the adaptive regularization scheme proposed in Sec.~\ref{sec:reg}, the entries of the regularization matrix would tend to infinity as the respective parameter approaches zero.
To prevent this, we limit the breadth of the parameter space by setting $|(\bm{A}^\text{norm})_{kl}| \geq 10^{-5}$ for all $k\neq l$ and accordingly set the step size for evaluating finite differences to $10^{-5}$.
Finally, as $t^{\text{norm}}\leq 1$, we limit $(\bm{A}^{\text{norm}}_{ss})_{kk} \geq 2$ for all $k$.
We do not set an upper boundary for any parameter. 
We enforce the bounds by setting a parameter to the minimal value whenever an update would otherwise fall below that value. 
\subsection{Initial guess}
\label{app:guess}
We can make a reasonable heuristic initial guess within the normalized parameter space described in Sec.~\ref{app:norm}.
Consider the case where the off-diagonal entries of $\bm{A}_{ss}$ are set to zero. 
Eq.~\ref{eq:G_to_A} then yields the integral of a sum of $h$ exponential,\cite{ceriottiNovelFrameworkEnhanced2010} where the time constant of the $k$th exponential is given by $\left(\bm{A}_{ss}\right)_{kk}^{-1}$.
We want to ensure that the initial guess is plateaued on the relevant time scale ($t\approx 1$) and, at the same time, can be sufficiently well resolved by the time grid. 
Therefore, we set $10\Delta t / t_{\text{max}} \leq \left(\bm{A}_{ss}\right)_{kk}^{-1} \leq 0.5$, or equivalently, $2\leq\left(\bm{A}_{ss}\right)_{kk}\leq t_{\text{max}}/ 10 \Delta t$.
More specifically, we choose $\left(\bm{A}_{ss}\right)_{kk}^{-1}$ to be logarithmically equispaced within these bounds.
The plateau value of the $k$th integrated exponential is given by $\left(\bm{A}_{ss}\right)_{kk}^{-1}\left(\bm{A}_{Ps}\right)_k^2$.
As an initial guess, all exponentials should equally contribute to the total (long time) friction.
Thus, to achieve a total plateau value of $\approx1$ we set $\left(\bm{A}_{Ps}\right)_k = \sqrt{\frac{\left(\bm{A}_{ss}\right)_{kk}}{h}}$. 
Finally, the off-diagonal entries of $\bm{A}_{ss}$ should not be set to zero for the initial guess, as the corresponding entries in the Jacobian would otherwise vanish. 
We set the off-diagonals to $(\bm{A}_{ss})_{kl}  = (\bm{A}_{ss})_{kk}/(l-k) $  for all $k<l$.
\section{Gauss-Newton method for fitting memory kernels}
\label{app:GN-fit}
How well a Gauss-Newton method performs can depend strongly on the specific application. Further strategies beyond those described in sections ~\ref{sec:reg} and \ref{app:step} may be necessary to improve reliability and convergence.
As an intermediate step in finding an optimized procedure for solving Eq.~\ref{eq:iomk_gn_loss}, we first solved the more straightforward problem of fitting a memory kernel $\tilde{G}(t)$ using Eq.~\ref{eq:G_to_A}.
To do this, we set $\bm{f}:= \bm{G}(\bm{A})$, $\bm{f}^{\text{tgt}}:=\tilde{\bm{G}}$ and $\bm{x} := \text{all distinct entries in } \bm{A}$ and evaluate the  Jacobian $\bm{J}$ by finite differences.
With these definitions, applying the Gauss-Newton method is, in principle, straightforward.
The strategies described in Appendix \ref{app:norm} and \ref{app:bounds} can be applied to achieve stable and efficient convergence in both IOMK-GN and memory kernel fitting. 
\subsection{Fitting exemplary integrated memory kernels using the Gauss-Newton method} \label{app:fit}
As the Jacobian for fitting
\begin{equation}
    J_{mn} = \left.\frac{\partial \tilde{G}^i_n}{\partial x_m}\right|_{\bm{A}^i}, 
\end{equation}
only differs from the IOMK-GN Jacobian (compare Eq.~\ref{eq:iomk-gn_jacobian})  by a time-dependent factor, we assume that we can infer a good choice for the regularization parameter $\alpha$ and compare the two regularization schemes for fitting as described in Sec.~\ref{app:GN-fit}.

\begin{figure}
    \centering
    \includegraphics{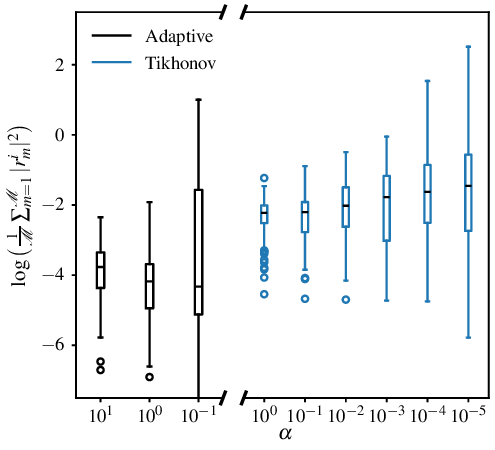}
    \caption{
    The statistics of the logarithm of the squared residuals averaged over all entries of the integrated memory kernel for the Tikhonov and adaptive regularization schemes, respectively, using different regularization parameters. The residuals are evaluated based on the normalized representation of the respective target.
    The box spans the distance between the first and third quartiles, with the median indicated by a black line.
    The whiskers indicate the largest and smallest data point found within $1.5$ box lengths outside the box. 
    All remaining points are considered outliers and are marked with small circles.
    For the sake of clarity, outliers or whiskers below $-7.5$, only present in the adaptive scheme with $\alpha=10^0$ and $\alpha=10^{-1}$ respectively, are not shown. The Tikhonov regularization with $\alpha=10^{-5}$ has outliers up to $\approx 5$.
    In the rare case that the fitting process failed due to non-invertible $\bm{D} + \boldsymbol{\Lambda}$ (only in $\alpha=10^{-1}$ in the adaptive scheme), we set the error to 10.0 (1 on the logarithmic scale).
    The data was obtained by fitting Eq.~\ref{eq:G_to_A} to 100 randomly generated integrated memory kernels. 
    }
    \label{fig:screen_fit}
\end{figure}

To that end, we fitted Eq.~\ref{eq:G_to_A} with $\bm{A}\in \mathbb{R}^{9\times 9}$ to 100 randomly generated memory kernels (see Appendix ~\ref{app:fit} for details) and show the mean squared error after 100 iterations for considered regularizations in  Fig.~\ref{fig:screen_fit}.
For the Tikhonov regularization, we found the best results for $\alpha = 0.01$, while the adaptive regularization scheme performs optimally for $\alpha = 1$. Independent of the choice of $\alpha$, the adaptive scheme yields, on average, errors that are orders of magnitude smaller than the Tikhonov scheme.

\section{$2\times 2$ block diagonal drift matrices}
\label{app:block}
Any exponential matrix of the form $e^{-\bm{A}_{ss}t}$ where $\bm{A}_{ss}\in\mathbf{R}^{h\times h}$ can be described as a sum of complex exponentials.
Diagonalizing $\bm{A}_{ss}$ with its eigenvector matrix $\bm{U}$ yields a eigenvalue matrix $\bm{\Gamma}$ such that
\begin{equation}
    \bm{A}_{ss} = \bm{U}\bm{\Gamma}\bm{U}^{-1}.
\end{equation}
This allows us to rewrite the exponential in Eq.~\ref{eq:G_to_A} as
\begin{equation}
    e^{-\bm{A}_{ss}t} = e^{-\bm{U}\bm{\Gamma}\bm{U}^{-1}t} = \bm{U}e^{-\bm{\Gamma}t}\bm{U}^{-1}.
    \label{eq:exp_diag}
\end{equation}
Eigenvalues of an invertible squared real matrix with even $h$ always come in conjugate pairs of the form
\begin{equation}
    \gamma_i^\pm = \sigma_i \pm \chi_i = \sigma_i \pm \sqrt{\kappa_i},
    \label{eq:eig}
\end{equation}
where $\sigma_i, \kappa_i \in \mathbb{R}$.
Additionally, we only consider $\sigma_i > 0$ such that the resulting memory decays for $t\rightarrow \infty$.
While the eigenvalues of $\bm{A}_{ss}$ may be complex, the codomain of Eq.~\ref{eq:exp_diag} is real.
Given $h_r$ pairs of real and $h_c$ pairs of complex eigenvalues Eq.~\ref{eq:K_to_A2} takes the general form
\begin{equation}
\begin{split}
    \tilde{K}(t) = &\sum_{i=1}^{h_r}\left( \alpha_i e^{-\gamma_i^+|t|} + \beta_ie^{-\gamma_i^-|t|}\right)\\
    &+\sum_{i=h_r+1}^{h/2} e^{-\sigma_i|t|}(\alpha_i \cos(\chi_i|t|)+\beta_i \sin(\chi_i|t|))\\
\end{split}
\label{eq:K_A_general}
\end{equation}
where $\alpha_i$ and $\beta_i$ are real coefficients.
Eq.~\ref{eq:K_A_general} suggests that in principle, $h$ real numbers suffice to characterize a matrix similar to any $\bm{A}_{ss}$, which with $h$ additional parameters, fully characterizes the memory kernel. 

For example, $A_{ss}$ can be constructed as a $2\times 2$ block-diagonal matrix such that~\cite{liComputingNonMarkovianCoarsegrained2017a}
\begin{equation}
    (\bm{A}_{ss})_{kk} = \begin{pmatrix}
        a_k & b_k\\
        - b_k & 0 
    \end{pmatrix}.
    \label{eq:22_block}
\end{equation}
When $h=2$, this yields a single $2\times 2$ block, and the full matrix can be written as

\begin{equation}
    \bm{A} = \begin{pmatrix}
           0 & c_1 & c_2\\
        -c_1 & a & b\\
        -c_2 &-b & 0 
    \end{pmatrix}.
    \label{eq:one_block}
\end{equation}
As indicated by Eq.~\ref{eq:eig}, this form allows to model memory kernels of a single damped oscillator corresponding to complex eigenvalues ($b>0.5a\rightarrow \kappa < 0$), two exponentials corresponding to real eigenvalues ($b\leq 0.5a\rightarrow \kappa \geq 0$), or a single exponential where one eigenvalue is zero ($b=0\rightarrow \sqrt{\kappa} = \sigma$). 
Note that directly fitting damped oscillators to a memory\cite{liComputingNonMarkovianCoarsegrained2017a,wangImplicitsolventCoarsegrainedModeling2019,klippensteinCrosscorrelationCorrectedFriction2021} only covers the case of $\kappa \leq 0$.

While in principle Eq.~\ref{eq:22_block} provides the full range of possible kernels, constructing $\bm{A}_{ss}$ as a block-diagonal matrix might be suboptimal for the optimization process.

To get an intuition, consider that the total contribution of one block to the integral of the memory kernel increases with $1/b_i^2$ (see Eq.~\ref{eq:G_to_A}).  
In the limit of $b_i\rightarrow 0$, the integral of the memory kernel diverges to infinity. 
Accordingly, partial derivatives with respect to $b_i$ diverge in this limit, while partial derivatives with respect to the other parameters tend to zero, which can simultaneously destabilize and slow down the convergence of the GN updates.

By adding a single additional parameter per $2\times 2$ block
\begin{equation}
    \bm{A} = \begin{pmatrix}
           0 & c_1 & c_2\\
        -c_1 & a & b\\
        -c_2 &-b & d 
    \end{pmatrix},
     \label{eq:one_block2}
\end{equation}
additional entries, specifically diagonal entries, enter the determinant of $\bm{A}_{ss}$, and the integrated memory kernel scales with
$1/(a_id_i+b_i^2)$.
We would expect that this reduces the chance that the determinant becomes too small and, therefore, should improve the optimization's numerical stability.
By extension, requiring all entries in $\bm{A}$ to be non-zero should eliminate diverging partial derivatives. 

As a numerical test, we fitted the random test memory kernels of Appendix~\ref{app:GN-fit}, using $2\times 2$ block diagonal matrices of the form of  Eq.~\ref{eq:one_block} and \ref{eq:one_block2} such that $\bm{A}\in \mathbb{R}^{9\times9}$, otherwise using the same bounds and regularization and a comparable initial guess.

\begin{figure}
    \centering
    \includegraphics{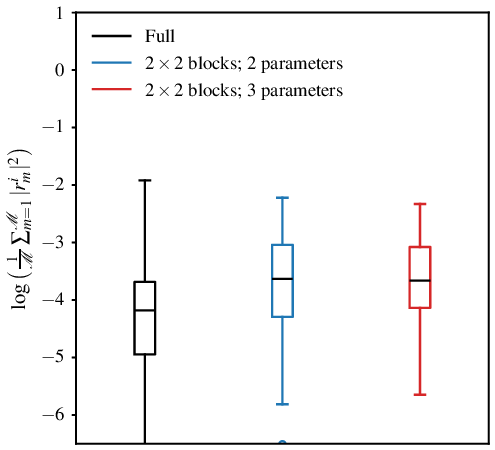}
    \caption{The statistics of the logarithm of the squared residuals averaged over 100 randomly generated memory kernels using the full matrix, a $2\times2$ block-diagonal matrix with $2$ and $3$ parameters per block, respectively.
    The residuals are evaluated as in, and the box representation follows Fig.~\ref{fig:screen_fit}, with the difference that 20/30 failed optimizations for 2 and 3 parameters, respectively, are excluded.}
    \label{fig:block_fit}
\end{figure}
While the median error is of the same order of magnitude for all three cases, it is smallest for the full matrix. 
For the 2-block matrices in $20\%$ / $30\%$ of cases, the optimization failed due to non-invertible $\bm{D}+\bm{\Lambda}$, which was not the case for the full matrix.
In conclusion, while for the tested cases, extending the $2\times 2$ blocks by a single additional parameter did not systematically improve the stability and quality of the fits, using the full matrix did.

\end{document}